\documentclass[twocolumn,apj,numberedappendix,twocolappendix,iop]{openjournal}

\usepackage[T1]{fontenc}
\usepackage{graphicx}	
\usepackage{amsmath}	
\usepackage{newtxtext,newtxmath}    
\usepackage[dvipsnames]{xcolor}  
\usepackage[colorlinks,linkcolor=blue,citecolor=blue,urlcolor=blue ]{hyperref}
\usepackage{amssymb}	
\usepackage{dsfont}
\usepackage{ulem}
\usepackage{soul}
\usepackage{enumitem}
\usepackage{orcidlink}
\usepackage{hyperref}

\usepackage{natbib}

\usepackage{fontawesome}
\usepackage{lineno}

\usepackage{macros} 

\begin{document}


\title{Universal numerical convergence criteria for subhalo tidal evolution}

\shorttitle{Numerical convergence of subhalo tidal evolution}
\shortauthors{Chiang et al.}

\author{Barry T. Chiang$^{1,2}$\orcidlink{0000-0002-9370-4490}}
\author{Frank C. van den Bosch$^1$\orcidlink{0000-0003-3236-2068}}   
\author{Hsi-Yu Schive$^{2,3,4,5}$\orcidlink{0000-0002-1249-279X}}

\affiliation{$^1$Department of Astronomy, Yale University, PO. Box 208101, New Haven, CT 06520-8101}
\affiliation{$^{2}$Institute of Astrophysics, National Taiwan University, Taipei 10617, Taiwan}
\affiliation{$^{3}$Department of Physics, National Taiwan University, Taipei 10617, Taiwan}
\affiliation{$^{4}$Center for Theoretical Physics, National Taiwan University, Taipei 10617, Taiwan}
\affiliation{$^{5}$Physics Division, National Center for Theoretical Sciences, Taipei 10617, Taiwan}

\email{barry.chiang@yale.edu}  

\label{firstpage}


\begin{abstract}
    Dark matter subhalos and satellite galaxies in state-of-the-art cosmological simulations still suffer from the ``overmerging'' problem, where inadequate force and/or mass resolution cause artificially enhanced tidal mass loss and premature disruption. Previous idealized simulations addressing this issue have been restricted to a small subset of the subhalo orbital parameter space, and all assumed subhalos to be isotropic. Here, we present the first extensive simulation suite that quantifies numerical convergence in the tidal evolution of anisotropic subhalos under varying numerical resolutions and orbits. We report a universal force resolution criterion: the subhalo's instantaneous tidal radius must always be resolved by at least 20 cells in adaptive mesh refinement (AMR)-based simulations, or by 20 softening lengths (Plummer equivalent) in tree-based simulations, regardless of refinement details or subhalo physical properties such as concentration or velocity anisotropy. We also report a universal expression for the discreteness-noise-driven scatter in the bound-mass fraction of subhalos that depends only on the subhalo mass resolution at infall and the instantaneous bound mass fraction, agnostic of any further subhalo properties. Such stochastic discreteness noise causes both premature disruption and, notably, spurious survival of poorly mass-resolved subhalos. We demonstrate that as many as 50 percent of all subhalos in state-of-the-art cosmological simulations are likely to be either force and/or mass unresolved. Our findings advocate for adaptive softening or grid refinement based on the instantaneous tidal radius of the subhalo.
\end{abstract}


\maketitle


\section{Introduction}
\label{sec:Introduction}

In the established paradigm of hierarchical structure formation \citep[e.g.][]{Peebles1971phcoP, Cole2000MNRAS319168C, Frenk2012AnP524507F, Bland-Hawthorn2016ARA&A54}, the growth of structure is driven by the continuous accretion of lower-mass systems. Once accreted, these ``substructures'' undergo tidal evolution that causes them to spiral inward due to dynamical friction, while experiencing mass loss due to tidal stripping and tidal shocking \citep[e.g.][]{Chandrasekhar1943ApJ97255C, Gnedin.etal.99, vandenBosch2018MNRAS4743043V}. 

The abundance and demographics of dark matter substructures provide important small-scale probes of the nature of dark matter via observables that include, among others, the decay signals of dark matter particles \citep[e.g.,][]{Strigari.etal.07, Pieri.etal.08,  Ando2019Galax768A}, gravitational lensing distortions \citep[e.g.,][]{Vegetti.etal.14, Hezaveh.etal.16, Meneghetti.etal.20}, gaps in stellar streams \citep[e.g.,][]{Carlberg.12, Erkal.etal.16, Bonaca.etal.19}, and the abundance of satellite galaxies \citep[e.g.,][]{Anderhalden.etal.13, Nadler.etal.21}. In order for the inference of these methods to yield unbiased constraints on the nature of dark matter, it is of crucial importance that we are able to make predictions for the substructure characteristics for a given dark matter model that are both accurate and precise.

Because of the non-linearities involved, the tidal evolution of dark matter substructure is typically studied using numerical simulations. However, due to the finite mass and force resolution, these simulation results are not always reliable \citep[e.g.][]{vandenBosch2017MNRAS468885V, vdBosch.Ogiya.18}. In particular, over the years it has become clear that there are two separate numerical problems at play: inadequate subhalo identification and overmerging.

Subhalo-finding algorithms often lose track of or simply fail to identify subhalos, especially near the center of the host halo \citep[e.g.][]{Behroozi2013ApJ762109B, Carlsten2020ApJ902124C, Mansfield2024ApJ970178M, Diemer2024MNRAS5333811D}. This issue is particularly prevalent with configuration-space-based subhalo finders such as \texttt{HFOF} \citep{Klypin1999ApJ516530K}, \textsc{SubFind} \citep{Springel2001MNRAS328726S}, \texttt{SKID} \citep{Stadel2001PhDT21S}, and \textsc{AHF} \citep{Knollmann2009ApJS182608K}, which rely only on the spatial clustering of particles. Several halo-finder comparison studies \citep[e.g.][]{Onions2012MNRAS4231200O, Knebe2013MNRAS4351618K, Behroozi2015MNRAS4543020B, vandenBosch2016MNRAS4582870V, ForouharMoreno2025arXiv250206932F} have demonstrated improved identification of subhalos when using the full 6D phase-space information (e.g., \texttt{6DFOF} \citep{Diemand2006ApJ6491D}, \textsc{Rockstar} \citep{Behroozi2013ApJ762109B}, \texttt{VELOCIraptor} \citep{Elahi2019PASA3621E}), or by tracking particles across simulation outputs (e.g., SURV \citep{Tormen.etal.98}, HBT+ \citep{Han2012MNRAS4272437H, Han2018MNRAS474604H}, \textsc{Sparta} \citep{Diemer2024MNRAS5333811D}, \textsc{Symfind} \citep{Mansfield2024ApJ970178M}, \texttt{Bloodhound} \citep{Kong2025arXiv250310766K}, \textsc{Haskap Pie} \citep{Barrow2025arXiv250522709B}). However, even with these more elaborate subhalo finders, there has been little sign of convergence in the radial distribution of subhalos \citep{Mansfield2024ApJ970178M, ForouharMoreno2025arXiv250206932F}.

Overmerging is a problem of the simulation itself. Whenever a subhalo is poorly resolved, it experiences too much mass loss, which causes the subhalo to dissolve prematurely \citep[e.g.][]{vandenBosch2018MNRAS4743043V, vdBosch.Ogiya.18, Amorisco2021arXiv211101148A, Errani2020MNRAS491, Errani2021MNRAS50518E, Martin2024MNRAS5352375M}. In fact, these and other studies \citep[][]{Drakos2020MNRAS494378D, Drakos2022MNRAS516106D, Benson2022MNRAS5171398B, Stucker2023MNRAS5214432S} have argued that a properly resolved isotropic cold dark matter (CDM) subhalo should never completely disrupt in the absence of baryons. It has been shown that this artificial disruption causes a significant underprediction of the substructure abundance with potential far-reaching ramifications for cosmological inference \citep[][]{vandenBosch2017MNRAS468885V, Webb2020MNRAS499116W, Green2021MNRAS5034075G, Grand2021MNRAS5074953G, Mansfield2021MNRAS5003309M, AguirreSantaella2023MNRAS51893A, Mansfield2024ApJ970178M}. It might also explain why state-of-the-art cosmological simulations still manifest a pronounced deficit of surviving subhalos and satellite galaxies within the inner $0.1$\textendash$0.3$ of the host halo's extent, relative to the observed radial distribution of substructures in both massive galaxy clusters \citep{Natarajan2009ApJ693970N, Natarajan2017MNRAS4681962N, Ragagnin2022AA665A16R} and Milky Way-scale environments \citep{Kelley2019MNRAS4874409K, Graus2019MNRAS4884585G, Carlsten2020ApJ902124C, Carlsten2022ApJ93347C, Grand2021MNRAS5074953G, Lovell:2021vmq}. 

Properly simulating the tidal evolution of substructure requires both sufficient \textit{mass} and \textit{force} resolution. Formally, a simulated (sub)halo is a coarse-grained discretization of its actual phase-space distribution function. With decreasing mass resolution (i.e., decreasing number of macro-particles used to represent the substructure), the resulting discreteness noise introduces Poisson fluctuations in the density profile and gravitational potential of the subhalo, which adversely impacts its dynamical evolution \citep[e.g.][]{Romeo2004MNRAS3541208R, Romeo2008ApJ6861R}. In particular, as shown in \citet{vdBosch.Ogiya.18}, insufficient mass resolution introduces stochasticity in the subhalo bound mass fractions.

The force resolution is a measure for the smallest spatial scale on which local density gradients are accurately captured in the computation of the gravitational acceleration field, which depends on the architecture used for the gravity solver. Current cosmological and astrophysical simulation codes fall largely into two categories depending on how the gravitational force between nearby particles is calculated. Some codes compute the gravitational force on each particle by interpolating the acceleration field calculated on a grid-based mesh, which is typically allowed to be adaptive. In this case, the force resolution is set by the local instantaneous minimum cell size, $\Delta x^\text{ins}_\text{min}$, which is set by the refinement criterion and/or the maximum level of refinement allowed. Examples of this type of Adaptive Mesh Refinement (AMR) codes include \textsc{Art} \citep{Kravtsov1997ApJS11173K}, \textsc{Flash} \citep{Fryxell2000ApJS131273F}, \textsc{Ramses} \citep{Teyssier2002A&A385337T}, \textsc{Enzo} \citep{Bryan2014ApJS21119B}, and \textsc{gamer}~\citep{Schive2010ApJS186457S, Schive2018MNRAS4814815S}.

Other codes rely on some kind of tree algorithm to speed up the computation of gravitational forces between particles \citep[e.g.,][]{Barnes.Hut.86, Dehnen2002JCoPh17927D}. On small scales, such codes adopt explicit force softening to reduce the accelerations between nearby particles that would otherwise give rise to artificial large-angle scattering. In this case, the force resolution of the simulation is determined by the gravitational softening length $\varepsilon$, which is typically identical for all particles in the simulation box, regardless of their local environment \citep[but see][for methods with adaptive softening]{Price2007MNRAS3741347P, Iannuzzi2011MNRAS4172846I, Hopkins2023MNRAS5255951H}. Examples of this type of simulation codes include \textsc{Gadget} \citep{Springel2021MNRAS5062871S}, \textsc{ChaNGa} \citep{Menon2015ComAC21M}, \textsc{GyrfalcOn} \citep{Dehnen2002JCoPh17927D}, \textsc{Arepo} \citep{Springel2010MNRAS401791S}, and \textsc{Gizmo} \citep{Hopkins2015MNRAS45053H}. 

Using a suite of idealized simulations, \citet{vdBosch.Ogiya.18} determined criteria under which subhalos in numerical simulations may be deemed force-resolved and mass-resolved. They showed that the uncertainty in the bound mass fraction due to discreteness noise (i.e., limiting mass resolution) starts to exceed 0.1~dex once the bound mass fraction of the subhalo, $f_{\rm bound}$, drops below a minimum value of $0.32 (N_{\rm par}/1000)^{-0.8}$, where $N_{\rm par}$ is the number of particles of the subhalo at accretion. Similarly, a subhalo becomes force-unresolved when $f_{\rm bound}$ drops below a critical value that scales linearly with the softening length used. Hence, not surprisingly, better resolving the tidal evolution of substructure requires simulations with a larger number of particles and with smaller softening. Importantly, there will {\it always} be a bound mass fraction below which a subhalo is either mass- or force-unresolved. In other words, numerical artifacts are unavoidable, and the only cure is to define detailed mass-resolution and force-resolution criteria that can be used to flag subhalos in the simulation box that are deemed unreliable.

Although the study of \citet{vdBosch.Ogiya.18} presented such resolution criteria, their analysis suffers from two important shortcomings. First, they are only valid for simulations that use tree-based algorithms with force-softened particles. To our knowledge, no study has yet investigated a criterion for the force resolution, $\Delta x^\text{ins}_\text{min}$, required to resolve the tidal evolution of dark matter subhalos in AMR-based simulations, despite the fact that many studies of subhalo demographics have been based on AMR-based simulations such as the Bolshoi simulation \citep{Klypin2011ApJ740102K}, the MultiDark simulations \citep{Prada2012MNRAS4233018P}, and the Horizon-AGN suite \citep{Dubois2014MNRAS4441453D}. 

Second, \citet{vdBosch.Ogiya.18} based their analysis of force- and mass-resolution criteria on idealized simulations of isotropic halos. Several studies have shown that such isotropic subhalos evolve along a universal ``tidal track,'' a tight correlation between the bound mass fraction and subhalo structural parameters (e.g. peak circular velocity), irrespective of the detailed host potential or the subhalo orbits \citep[e.g.][]{Penarrubia2010MNRAS4061290P, Green2019MNRAS4902091G, Errani2021MNRAS50518E, Benson2022MNRAS5171398B}. However, cosmological simulations have shown that the velocity structure of dark matter halos clearly deviates from isotropy \citep[e.g.,][]{Diemand.etal.07, Navarro2010MNRAS40221N, Ludlow2011MNRAS4153895L, Klypin2016MNRAS4574340K, He2024arXiv240714827H}. More importantly, \citet{Chiang2024arXiv241103192C} have shown that velocity anisotropy strongly impacts tidal evolution, with subhalos that are radially anisotropic experiencing much more mass loss than their tangentially anisotropic counterparts. In fact, subhalos with different velocity anisotropies evolve along different tidal tracks. This implies that they have a different density structure when stripped by the same amount, even when they started out with the same density profile prior to infall. Hence, it may well be that the force resolution criterion depends on the velocity anisotropy of a subhalo.

In order to address these shortcomings, we present the first comprehensive study of numerical convergence and resolution-limited artifacts in AMR codes of the tidal evolution of subhalos with different velocity anisotropies. We analyze a large suite of idealized simulations run with the AMR code \gamer~\citep{Schive2018MNRAS4814815S}, and demonstrate the existence of a simple universal force resolution criterion: {\it a subhalo is ``force-resolved'' if its instantaneous tidal radius is always resolved by at least 20 cells}. This criterion is independent of internal subhalo properties, such as concentration or velocity anisotropy. We also show that the numerical scatter in the bound mass fractions of subhalos due to discreteness noise follows a universal evolution track that is agnostic of numerical refinement details or physical subhalo attributes. We show that the subhalo resolution criteria obtained here for AMR-based codes are in agreement with the criteria of \citet{vdBosch.Ogiya.18} for tree-based simulations; both suggest that the optimal method for resolving the tidal evolution of subhalos is to adopt a tidal-radius-based adaptive softening or refinement scheme in future simulations. 

Finally, we apply our force resolution criteria to a catalog of subhalos selected from the Bolshoi simulation \citep{Klypin2011ApJ740102K} and find that $\simeq50\%$ $(40\%)$ of all subhalos with more than 50 (200) particles are force-unresolved. This underscores that subhalo demographics inferred from state-of-the-art cosmological simulations are still subject to large uncertainties. We hope that the criteria presented here will facilitate more precise results by
restricting the analysis to a smaller sample of well-resolved subhalos.

The paper is organized as follows. \S\ref{sec:Simulation_Setup} describes the simulation setup. \S\ref{sec:Case_Study} presents a case study used to quantify the numerical convergence of subhalo properties under varying mass and force resolutions. In \S\ref{sec:Universal_AMR_Force}, we report the tidal-radius-based force resolution criterion for subhalos in AMR-based simulations, demonstrating its universality against numerical mass resolution (\S\ref{ssec:Universal_AMR_Force_Mass}), mesh refinement strategies (\S\ref{ssec:Universal_AMR_Strategies}), and physical properties of subhalos (\S\ref{ssec:Universal_AMR_Force_Sizes}). In \S\ref{sec:Numerical_Convergence}, we quantify the realization-to-realization scatter in the bound mass fractions of subhalo at different mass resolutions. \S\ref{sec:Current_Literature} discusses the equivalence of resolution criteria in AMR- and tree-based codes and shows that a large fraction of subhalos in typical AMR-based cosmological simulations are ``force-unresolved'' (i.e., in violation of our force resolution requirement) and therefore unreliable. Finally, \S\ref{sec:Conclusions} summarizes our conclusions. Throughout this paper, we follow the setup of \citet{Chiang2024arXiv241103192C} and adopt $H_0 = 70$ km s$^{-1}$Mpc$^{-1}$, giving a Hubble time of $\tH \eee H_0^{-1} = 13.97$~Gyr. We use $\ln$ and $\log$ to indicate the natural and 10-base logarithms, respectively.


\section{Methodology}
\label{sec:Simulation_Setup}

\subsection{Initial conditions and subhalo orbits}
\label{ssec:Simulation_Setup}

We evolve individual dark matter (sub)halos orbiting in a static NFW background potential (i.e. the ``host'' halo), following the overall numerical setup of \citet{Chiang2024arXiv241103192C}. All subhalos are sampled from spherically symmetric distribution functions that, before truncation, correspond to the NFW profile \citep{Navarro:1996gj}
\begin{align}\label{eqn:rho_NFW}
	\rho(r) = \rho_0\bigg(\frac{r}{\rsz}\bigg)^{-1}\bigg(1+\frac{r}{\rsz}\bigg)^{-2},
\end{align}
where $\rsz$ denotes the initial subhalo scale radius. The (sub)halo concentration is $c \eee \rvir/\rsz$, where the virial radius $\rvir$ is defined as the radius within which the average density is $\Delta_\text{vir} = 97$ times the critical density \citep{Bryan1998ApJ49580B}. The subhalo virial mass $\msz$ denotes the total mass enclosed within $\rvir$. Throughout the work, we fix the host halo concentration $c_\text{h} = 5$. For a subhalo on a circular orbit of radius  $r_\text{cir} = 0.4r_\text{vir,h}$, the corresponding orbital period is $\torb = 4.8$~Gyr. 

We explore subhalos with the following ranges of initial characteristics: concentration parameter $4 \leq c \leq 40$, virial mass $-2.5 \leq \log(\msz/M_\rmh) \leq -4$, particle resolution $10^3 \leq \Npar \leq 5\times10^7$, and (radius-independent) velocity anisotropy $-0.5 \leq \beta \leq +0.5$. Here,
\begin{equation}\label{betadef}
\beta \equiv 1 - \frac{\sigma^2_\rmt}{2\sigma^2_\rmr}\,,
\end{equation}
relates the tangential $\sigma_\rmt$ and radial $\sigma_\rmr$ velocity dispersions. At our fiducial choice of $c = 10$ and $\msz = 10^{-3} \Mhalo$, we have $\rvir = 0.1\rvirh$. All subhalo initial conditions are constructed using the new, open-source Python package \texttt{PIANISTpy} (\textbf{P}article \textbf{I}nitial conditions with \textbf{ANIS}o\textbf{T}ropy) described in Chiang et al., (in prep). Importantly, \texttt{PIANISTpy} performs fully self-consistent truncation in phase space such that the particle initial conditions automatically have finite spatial extents and require no further relaxation in isolation. Each initial subhalo is truncated so that its total mass equals the virial mass of the untruncated subhalo.

The subhalo orbits are uniquely characterized by the orbital energy, $E_{\rm orb}$, and the orbital angular momentum, $L_{\rm orb}$. Equivalently, we instead characterize the orbits using the dimensionless orbital radius
\begin{align}
\bigRE \equiv r_{\rm circ}(E_{\rm orb})/\rvirh,
\end{align}
and the orbital eccentricity
\begin{align}
e \equiv \frac{\rapo-\rperi}{\rapo + \rperi}.
\end{align}
Here $r_{\rm circ}(E_{\rm orb})$ denotes the radius of a circular orbit of energy $E_{\rm orb}$, $\rvirh$ is the virial radius of the host halo, and $\rapo$ and $\rperi$ are the apo- and peri-centers of the orbit, which are the roots for $r$ of
\begin{align}\label{apoperi}
\frac{1}{r^2} + \frac{2[\Phi_\rmh(r)-E_{\rm orb}]}{L_{\rm orb}^2} = 0\,,
\end{align}
with $\Phi_\text{h}$ the gravitational potential of the host halo \citep[e.g.][]{vandenBosch1999ApJ51550V}. In this study, we consider orbits that cover the entire ranges $\bigRE \in (0,1]$ and $e \in [0,1)$.

\subsection{Numerical simulations}
\label{ssec:numsim}

All simulations are carried out using the code $\gamer$ \citep{Schive2018MNRAS4814815S}, which supports hybrid CPU/GPU parallelization and AMR with octree data structure. Following \citet{Chiang2024arXiv241103192C}, we adopt a simulation box of size $(8\rvirh)^3$, covered by a $128^3$ root grid (level zero). The large box size ensures that tidally stripped particles always remain a sufficient distance from the simulation boundaries, which have outflow boundary conditions. A cell is refined whenever it contains more than $N_\text{par/cell}$ dark matter particles, and by design, the refinement levels of adjacent patches can vary by at most one level. For our fiducial AMR strategy, we adopt $N_\text{par/cell} = 4$ and allow up to nine levels of refinement. For our fiducial subhalo, which has $\rvir = 0.1 \rvirh$ and $c = 10$, this translates to a minimum cell size (and thus a maximum spatial resolution) of $\Delta x_\text{min} = 0.0012\rvir = 0.012\rsz$. In \S~\ref{sec:Universal_AMR_Force}, we discuss a variety of convergence tests by varying both the maximum refinement level (which controls $\Delta x_\text{min}$) and $N_\text{par/cell}$. 

It is important to distinguish between the minimum cell size that {\it can} be achieved, $\Delta x_\text{min}$, which is set by the maximum refinement level specified by the user, and the minimum cell size that is achieved instantaneously, $\Delta x^\text{ins}_\text{min}$.  In general, $\Delta x^\text{ins}_\text{min}\geq \Delta x_\text{min}$, and equality will only be achieved if the structure in question has a sufficiently high density and/or if $N_\text{par/cell}$ is sufficiently small. Typically, a smaller value for $N_\text{par/cell}$ results in a more aggressively refined grid structure, which is more likely to reach the maximum allowed refinement level for a given density structure.

Regarding force calculations, particles are evolved on the standard ``kick-drift-kick'' (KDK) scheme. Mass density and force interpolation is carried out using the Triangular-Shape Cloud (TSC) scheme \citep{Hockney1988csupbookH}, which provides greater numerical accuracy at higher computational costs, compared to the standard Cloud-in-Cell (CIC) approach \citep[e.g.][]{Li2012JCAP01051L}.  

At each integration time step, the instantaneous center of mass (CoM) position and velocity of the subhalo are computed on-the-fly from the $5\%$ most bound particles, following the iterative procedure outlined in Section~2.3 of \citet{vdBosch.Ogiya.18}. During each iteration, the subhalo self-potential is first computed from the instantaneous bound dark matter particles. Next, each individual dark matter particle's total energy, defined with respect to the subhalo's CoM, is updated and sorted, after which the new CoM position and velocity are determined using the $5\%$ most bound particles. This process is iterated until the changes in CoM position and velocity are smaller than $10^{-4}\rvir$ and $10^{-4}V_\text{vir}$, respectively. The bound mass fraction is defined as
\begin{align}\label{eqn:f_bound}
\fbound(t) \equiv \frac{m_\text{sub}(t)}{\msz}=\frac{N_{\rm bound}}{\Npar},
\end{align}
where $m_\text{sub}$ is the instantaneous bound mass of the subhalo at time $t$, and $N_{\rm bound}$ denotes the number of instantaneous bound particles. Operationally, we define subhalo disruption as the epoch when $N_\text{bound}$ drops below $10$. We evolve subhalos flagged for disruption in this way for another $0.5\torb$ to protect against spurious fluctuations in $N_{\rm bound}$, resulting in a robust confirmation of disruption. In addition to the bound mass fraction, at each timestep, we also compute for each subhalo the mean velocity anisotropy $\beta_{50\%}$ of the $50\%$ most bound particles.

\section{Numerical Convergence: A Case Study}
\label{sec:Case_Study}

\begin{figure*}
	\includegraphics[width=0.99\linewidth]{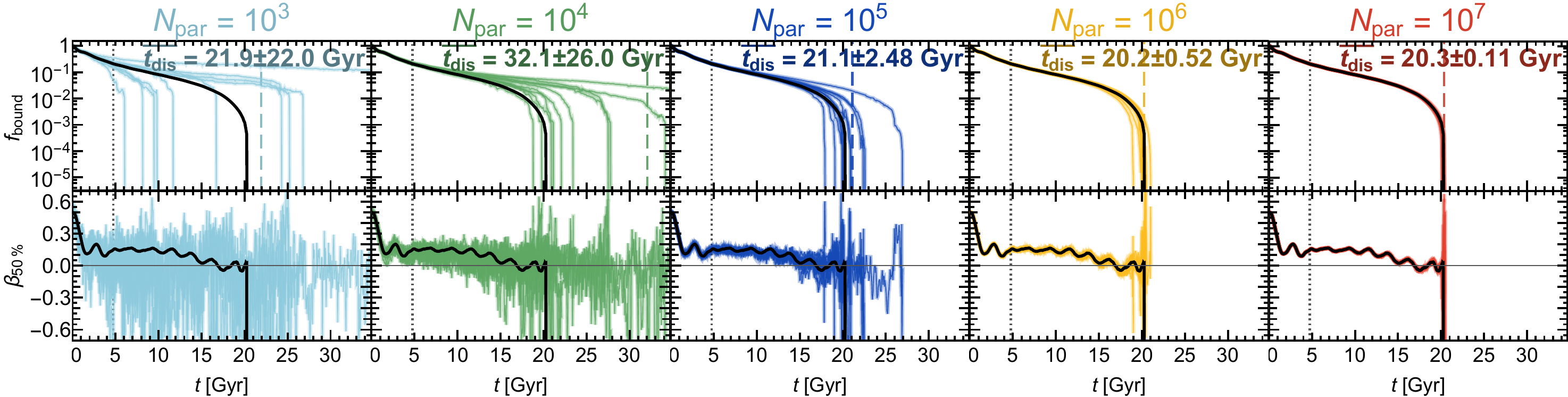}
	\caption{Evolution of the bound mass fraction $\fbound$ (top row) and mean velocity anisotropy $\beta_{50\%}$ (bottom row) of subhalos with an initial velocity anisotropy $\beta = 0.5$ along circular orbits with $\bigRE = 0.4$ (orbital period indicated by vertical dotted lines). Different columns correspond to different mass resolutions (i.e., different $\Npar$), as indicated at the top. Colored lines in each panel show the results for ten independent random realizations and are compared against the benchmark result (thick black curve) obtained using a simulation with $\Npar = 5 \times 10^7$. The ten-realization-averaged disruption time $t_\text{dis}$ (colored vertical dashed lines) and the associated one-sigma scatter are denoted in the top-right corner of each column. Note that the realization-to-realization variance becomes appreciable for $\Npar \lesssim 10^5$.} 
	\label{fig:Mass_Resolution}
\end{figure*}

This work seeks to establish the necessary and sufficient conditions to guarantee numerical convergence in the tidal evolution of anisotropic subhalos, particularly from the following three aspects in AMR-based simulations:
\begin{enumerate}[leftmargin=0.27truecm, labelwidth=0.2truecm]\renewcommand\labelenumi{\bfseries\theenumi}

\item \textbf{Mass resolution $\Npar$:} How does $\Npar$ correlate with the realization-to-realization scatter in the tidal evolution of subhalos driven by numerical discreteness noise? Does this correlation change under (in)adequate force resolution?
    
\item \textbf{Force resolution $\Delta x^\text{ins}_\text{min}$:} What is the minimum refinement grid size $\Delta x^\text{ins}_\text{min}$ required to guarantee convergent tidal evolution, above which a subhalo becomes ``force-unresolved'' (i.e. experiencing artificially inflated mass loss)? Is this threshold universal against varying subhalo physical properties?
    
\item \textbf{AMR strategies $N_\text{par/cell}$:} How do refinement strategies and criteria impact numerical convergence in AMR-based simulations?

\end{enumerate}

As a starting point, we use the fiducial resolution setup of \citet{Chiang2024arXiv241103192C}, which adopts $N_\text{par/cell} = 4$ and allows up to 9 levels of grid refinement, giving rise to a force resolution of $\Delta x_\text{min} = 0.0012\rvir$. We consider a subhalo of mass $m_{\rms,0}=10^{-3} M_\rmh$ with concentration $c=10$ and constant velocity anisotropy of $\beta = +0.5$, which is evolved along a circular orbit with $\bigRE=0.4$ in a host halo with concentration $c_\rmh = 5$. As shown in \citet{Chiang2024arXiv241103192C}, such a subhalo experiences pronounced anisotropy-driven core formation, which ultimately leads to complete physical disruption. We will use this setup as a useful benchmark to probe how different resolutions and AMR strategies impact the disruption time. Here, and throughout this paper, we assume that the results obtained with a simulation with $\Npar = 5\times 10^7$ represent the unbiased ``ground truth,'' against which we benchmark other simulations. We justify this choice in \fref{fig:Numerical_Convergence_Npar}. We consider subhalos to be ``force-resolved'' if the $\fbound(t)$ evolution, ensemble averaged over ten random realizations, converges correctly to $f^\text{truth}_\text{bound}(t)$ with unbiased realization-to-realization scatter.

\subsection{Mass resolution}
\label{ssec:Case_Study_Mass}

\begin{figure*}
	\includegraphics[width=\linewidth]{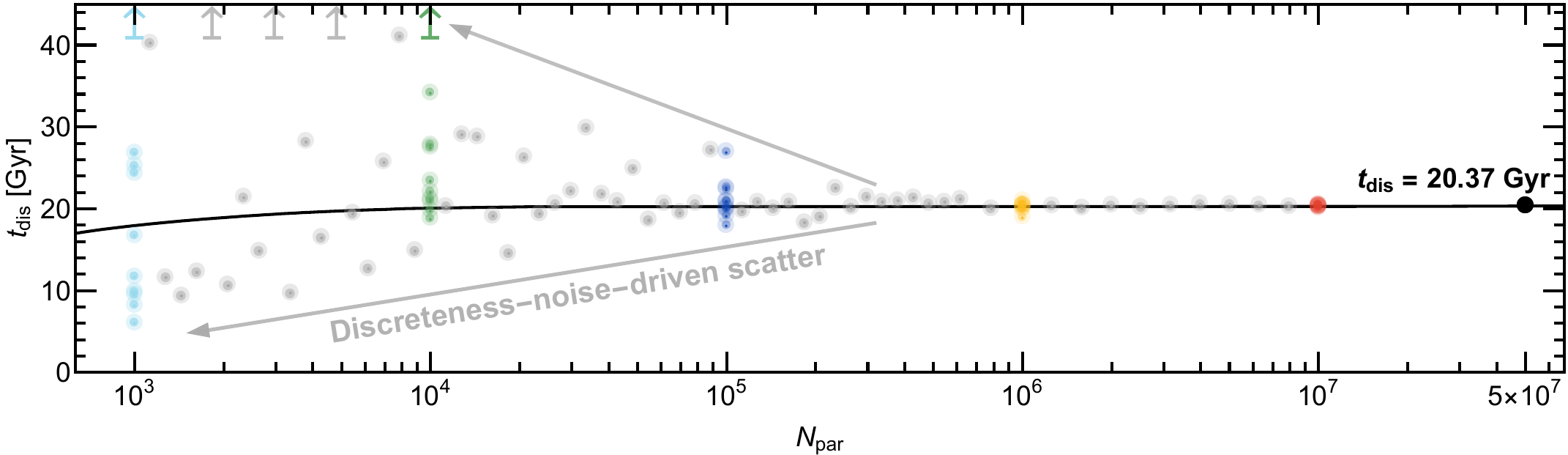}
	\caption{Numerical convergence of the tidal disruption time $t_\text{dis}$ of $\beta = 0.5$ subhalos evolved on $\bigRE = 0.4$ circular orbits with varying mass resolutions $\Npar = 10^3\text{\textendash}5\times10^7$. Colored dots are simulation results from \fref{fig:Mass_Resolution}, while the light-gray dots correspond to results based on additional simulations. Upward arrows indicate that the subhalos survive for longer than 45 Gyr. All simulations are based on a refinement scheme that assures adequate force resolution (based on the criteria discussed in \S\ref{ssec:Case_Study_Force}). The black curve indicates the $\Npar$-dependent disruption time defined as the time when a subhalo of $\Npar$ particles that follows $f^\text{truth}_\text{bound}(t)$ drops below 10 particles. Hence, in the absence of discreteness-noise-induced scatter, all data points should fall along this line. As is evident, discreteness noise becomes appreciable for $\Npar \lesssim 10^5$ and can cause both premature disruption (subhalo disrupts prior to the expected disruption time) as well as spurious survival (subhalo disrupts after the expected disruption time).}
	\label{fig:Numerical_Convergence_Npar}
\end{figure*}

\fref{fig:Mass_Resolution} compares the bound mass fraction $\fbound$ (top panels) and mean velocity anisotropy $\beta_{50\%}$ (bottom panels) evolution of subhalos resolved with $\Npar = 10^3$ (cyan), $10^4$ (green), $10^5$ (blue), $10^6$ (yellow), $10^7$ (red) equal-mass particles, compared against the $5\times10^7$ ``ground truth.'' We also indicate in each case, the average core disruption time $\overline{t_\text{dis}}$ (colored vertical dashed lines) and the associated one-sigma scatter, obtained from the ten random realizations.

Based on the ensemble-averaged $\fbound$ and core disruption time $\overline{t_\text{dis}}$, we infer that for $N_{\rm par} \leq 10^4$ the results are completely unreliable, with order unity simulation-to-simulation variance in the disruption times. With $N_{\rm par}=10^5$, this variance is reduced to $\sim 10\%$, and the mean evolution becomes comparable to that of the  ``truth,'' depicted by the solid black line. For $N_{\rm par} \geq 10^6$, the $\fbound$ evolution tracks are statistically consistent with the ``truth'' with negligible scatter down to $\fbound \lesssim 10^{-2}$ and the average disruption times are in good agreement with the value $t_\text{dis} = 20.37$~Gyr, obtained for the ``ground truth'' simulation with $\Npar = 5 \times 10^7$. A qualitatively similar trend is observed in the evolution of $\beta_{50\%}$. For $N_{\rm par} \leq 10^4$, the velocity anisotropies of the subhalos experience large fluctuations due to discreteness noise and do not follow a well-defined evolutionary track, while for $N_{\rm par} \geq 10^6$ the results are in good agreement with the ``truth,'' except for relatively large temporal fluctuations just prior to disruption, as the number of bound particles becomes too small for a reliable measure of the velocity anisotropy.

To better visualize the impact of discreteness noise, \fref{fig:Numerical_Convergence_Npar} plots the tidal disruption time as a function of $\Npar$ from a large suite of individual simulations, all for the fiducial setting but with different mass resolution. In addition to the simulations whose results are shown in \fref{fig:Mass_Resolution}, which are shown in color, we also run a set of simulations with different $\Npar$ that cover the entire range $10^3 \leq \Npar \leq 10^7$. As is evident, the disruption time starts to converge only after reaching $\Npar \gta 3 \times 10^5$, while subhalos below $\Npar < 10^5$ yield unreliable results, exhibiting numerical realization-to-realization scatter that increases rapidly as $\Npar$ decreases.

Note that while several subhalos with inadequate mass resolution disrupt much faster than the expected $t_\text{dis}$ inferred from the ``truth,'' others remain bound and survive for significantly longer periods of time. As a numerical artifact, these long-lived realizations consistently experience reduced tidal mass loss relative to the ``truth,'' leading to delayed disruption. In particular, the longest-surviving $\Npar = 10^3$~$(10^4)$ subhalo among the ten realizations survives until $t_\text{dis} = 80.7$~(104.8~Gyr). Hence, inadequate mass resolution does not cause a systematic error in the bound mass fractions, but rather causes a large statistical error. Overall, under adequate force resolution, decreasing mass resolution causes \textit{unbiased} widening in the discreteness-noise-driven scatter, with individual low-$\Npar$ realizations experiencing either a  net increase or a net decrease in their tidal mass loss leading to premature or delayed subhalo disruption, respectively.


\subsection{Force resolution}
\label{ssec:Case_Study_Force}

\begin{figure}
	\includegraphics[width=\linewidth]{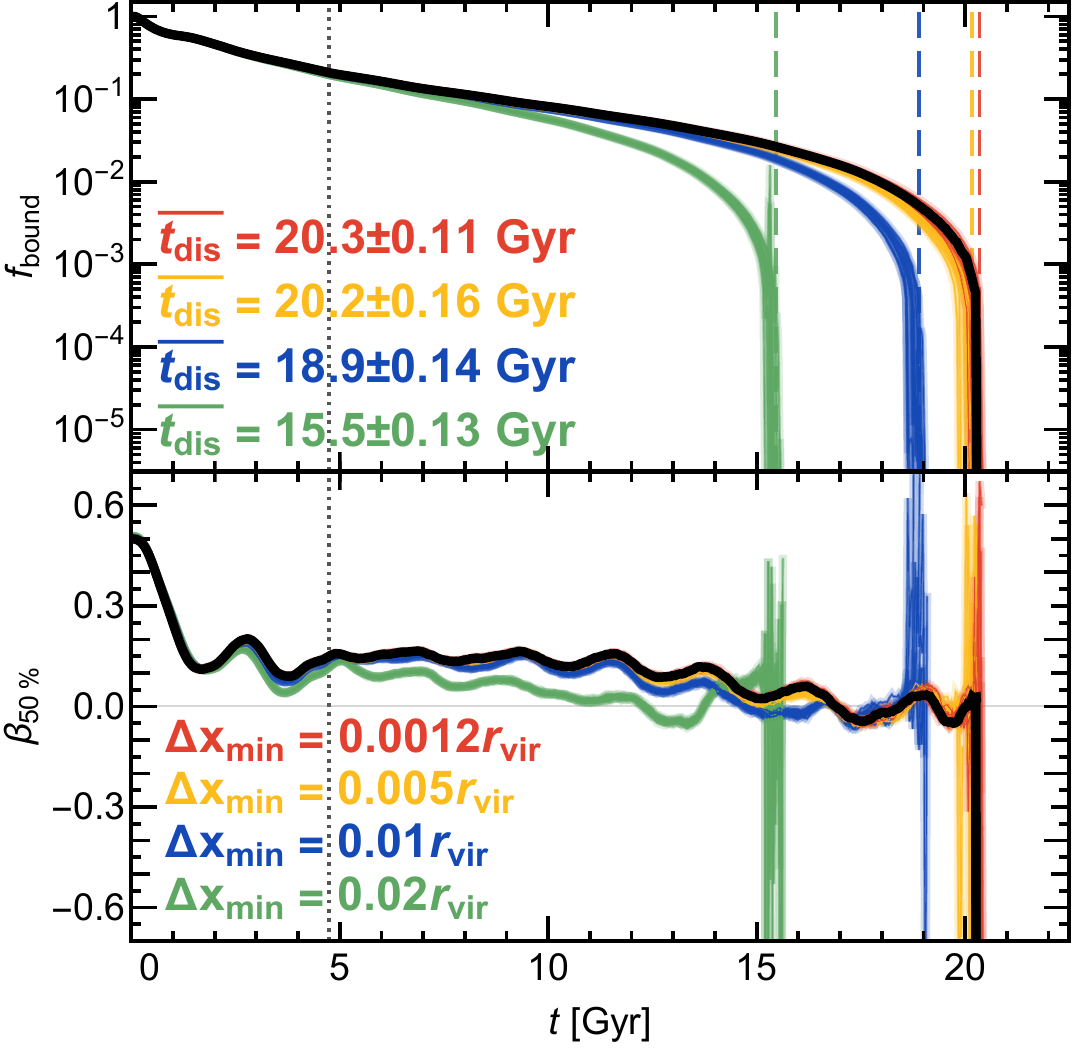}
	\caption{Evolution of $\fbound$ (top) and $\beta_{50\%}$ (bottom) for $\beta = 0.5$ subhalos evolved along circular orbits with $\bigRE = 0.4$, using different maximum-allowed refinement levels (color-coded as indicated). For each choice of $\Delta x_\text{min}$, we show the results for ten random realizations with $\Npar = 10^7$. The black solid lines indicates the ``truth'' obtained using a simulation with $\Npar = 5\times 10^7$ (black) and the fiducial $\Delta x_\text{min} = 0.0012\rvir$. Note that results are converged for $\Delta x_\text{min} \lesssim 0.01\rvir$. For $\Delta x_\text{min} = 0.02\rvir$, the subhalos are force-unresolved, giving rise to bound mass fractions that are too small which ultimately causes premature disruption. The vertical dotted line indicates the orbital period, while the vertical dashed lines mark the average disruption times.}
\label{fig:Force_Resolution}
\end{figure}

In order to investigate the minimum cell resolution scale $\Delta x^\text{ins}_\text{min}$ required for a subhalo to be considered force-resolved in AMR-based simulations, we take the set of ten $\Npar = 10^7$ subhalos with $\beta=+0.5$ used above and re-simulate them at progressively worse force resolutions. \fref{fig:Force_Resolution} compares the evolution of the subhalo bound mass fraction $\fbound$ (top panel) and the mean velocity anisotropy $\beta_{50\%}$ (bottom panel) obtained with different maximum-allowed refinement levels of 9 (red), 7 (yellow), 6 (blue), and 5 (green).  As indicated in the bottom panel, these correspond to minimum cell resolution scales ranging from $\Delta x_{\rm min} = 0.0012 \rvir$ for the fiducial refinement to level 9 to $0.02 \rvir$ when refinement is only allowed to level 5. Note that in each simulation, the minimum cell size achieved instantaneously, which we refer to as $\Delta x^\text{min}_\text{ins}$, is always equal to $\Delta x_{\rm min}$, at all times.

Note how, at a force resolution four times worse than the fiducial, all ten realizations still produce near perfect numerical convergence with negligible scatter. At $\Delta x_\text{min} = 0.01\rvir$, the subhalos disrupt prematurely, with the average disruption time, $\overline{t_\text{dis}}$, being underestimated by $\sim 7\%$. Note that the evolution in the velocity anisotropy is actually in remarkably good agreement with the truth all the way up to the point of disruption. At even worse force resolution, the ensemble-averaged evolution of both $\fbound$ and $\beta_{50\%}$ deviates strongly from the ``truth.'' Note that force-unresolved subhalos all experience artificially enhanced mass loss leading to premature tidal disruption and that the numerical realization-to-realization scatter appears insensitive to the adopted force resolution. Hence, poor force resolution only causes premature artificial disruption, but does not inflate numerical scatter at a fixed mass resolution. As we demonstrate in \S\ref{sec:Universal_AMR_Force}, this statement holds irrespective of the physical properties of the subhalos, and such numerical scatter can be reformulated as a universal scaling relation that depends only on $\Npar$ and $\fbound$ (\S\ref{sec:Numerical_Convergence}). These findings are in excellent agreement with those of \citet{vdBosch.Ogiya.18}, who showed that inadequate force softening in tree-based simulations (equivalent to inadequate cell refinement in AMR-based simulations) results in premature disruption, while poor mass resolution gives rise to large realization-to-realization scatter.
\begin{figure*}
	\includegraphics[width=0.99\linewidth]{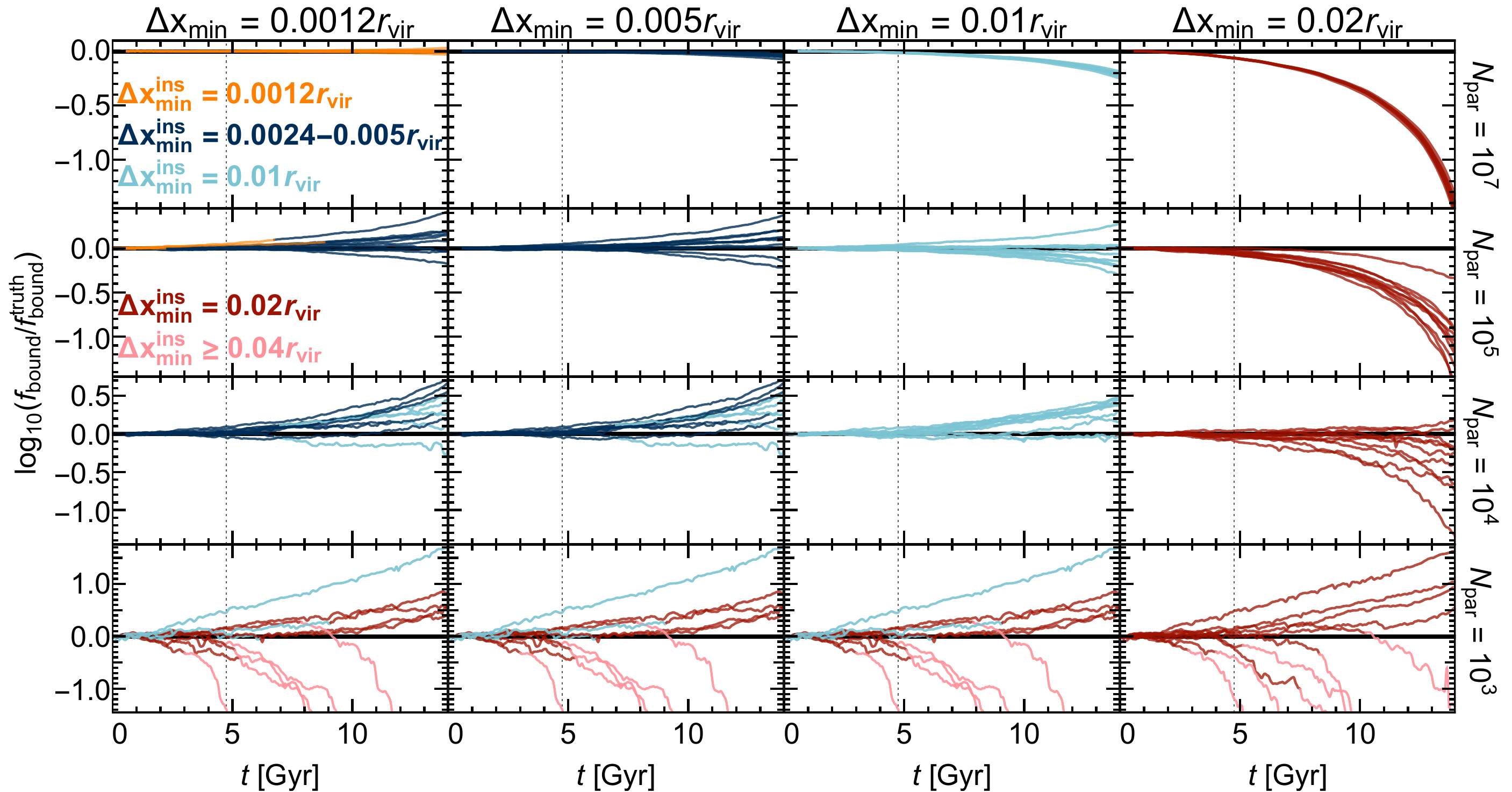}
    \caption{Evolution of the bound mass fractions, $\fbound$, normalized by $f^\text{truth}_\text{bound}$, of subhalos with $\beta=0.5$ evolved along circular orbits with $\bigRE=0.4$. 
    Different rows correspond to simulations with different mass resolution, ranging from $\Npar = 10^7$ to $10^3$, as indicate at the right-hand side. Different columns correspond to different maximum-allowed refinement levels of, from left to right, 9, 7, 6 and 5, resulting in the minimum cell sizes indicated at the top. Each panel shows the results for ten independent realizations that are individually color-coded based on the highest refinement level that is instantaneously achieved; 9 (orange), 8\textendash7~(dark blue), 6~(light blue), 5~(dark red), or 4 and below (pink). With $\Npar = 10^7$ the target refinement level is always achieved (i.e. $\Delta x^\text{ins}_\text{min} = \Delta x_\text{min}$). However, at lower mass resolution the instantaneous refinement level is lower than the target at late times due to the decreasing central particle number density. Note how force-unresolved subhalos (those with $\Delta x^\text{ins}_\text{min} > 0.01\rvir$, indicated as dark red or pink) typically experience artificially enhanced tidal mass loss, ultimately resulting in premature numerical disruption. However, if $\Npar$ is sufficiently small, the discreteness noise becomes sizable enough to cause some of the force-unresolved systems to have bound mass fractions that are artificially high.}
	\label{fig:Force_Conv_MassRes}
\end{figure*}

Based on the results presented so far, in what follows we consider a subhalo force-resolved if the instantaneous refinement reaches down to $\Delta x^\text{ins}_\text{min} \leq 0.01\rvir$. However, we will derive a more general force resolution criterion in \S\ref{ssec:Universal_AMR_Force_Sizes}.


\section{Universal AMR Force Resolution Criterion}
\label{sec:Universal_AMR_Force}

\subsection{Universality against mass resolution}
\label{ssec:Universal_AMR_Force_Mass}

\fref{fig:Force_Conv_MassRes} examines the numerical convergence of the bound mass fraction in our fiducial $\beta = 0.5$ subhalos simulated at different mass resolutions of $\Npar = 10^7, 10^5, 10^4, 10^3$ (top to bottom rows) and with maximum grid refinement up to levels~9, 7, 6, and 5 (left to right columns). The corresponding minimum cell resolution scales $\Delta x_{\rm min}$ are indicated at the top. Each panel shows the results for ten random realizations, with the color indicating the highest level of refinement achieved {\it instantaneously}. Note that the latter can change with time. Typically, a subhalo will start out with $\Delta x^\text{ins}_\text{min} = \Delta x_{\rm min}$, but as time progresses, and the central density of the subhalo decreases due to the tidal core-creation, occasionally the instantaneous level of maximum refinement will drop below the maximum allowed level. 

With $\Npar = 10^7$, the maximum allowed refinement level is achieved, that is, $\Delta x^\text{ins}_\text{min} = \Delta x_\text{min}$, at all times in all ten realizations. Note, though, that when the subhalos become force-unresolved (rightmost columns), they experience enhanced mass loss. With $\Npar = 10^5$, one starts to notice an appreciable scatter in the bound mass fractions across realizations, but as long as $\Delta x_{\rm min} \lesssim 0.01 \rvir$, the simulations are properly force-resolved, and the average $\fbound(t)$ is in good agreement with that of the ``truth.'' Note that with $\Delta x_\text{min} = 0.0012\rvir$ (leftmost panel), the refinement does not always reach the maximum allowed level. This simply reflects that the mass resolution is too low to 
resolve particle number densities that require a refinement to better than level 7. However, as long as the simulation is force resolved, this lack of mass resolution does not affect the {\it average} mass evolution of the subhalo. The results are qualitatively similar for $\Npar = 10^4$; poor mass resolution causes enhanced realization-to-realization scatter, while inadequate force resolution results in enhanced mass loss. 

Lastly, with $\Npar = 10^3$ the mass resolution is so poor that the realization-to-realization variance completely dominates; in particular, the results for the highest force resolution case are almost indistinguishable from those for the lowest force resolution. In fact, in the latter case, although the simulations are force-unresolved, in some cases the subshalos experience a {\it reduced} mass loss compared to the ``truth.'' This is rather surprising in light of the fact that simulations with a mass resolution that is four orders of magnitude higher always result in artificial, premature disruption. Based on the colors of the lines in the lower panels of \fref{fig:Force_Conv_MassRes}, it is clear that with $\Npar=10^3$ the number density of the subhalo is so low that it cannot reach or maintain $\Delta x^\text{ins}_\text{min} \leq 0.01\rvir$, which implies that the system is formally force-unresolved. However, because the mass resolution is so low, it does not behave like a force-unresolved system. Hence, in any cosmological simulation one should always expect to find some low-mass subhalos that have ``spuriously'' survived in the sense that they would have disrupted, or at least experienced more mass loss, if the simulation would have been run with higher mass resolution.

In summary, when a subhalo is poorly mass resolved ($\Npar \lesssim 10^3$), the results become completely unreliable, regardless of the force resolution used. When a subhalo is force-unresolved, it always experiences excessive mass loss (leading to premature or artificial disruption), except when the mass resolution is also extremely poor, in which case the discreteness-noise-dominated subhalo can spuriously survive and even experience a mass loss rate that is artificially suppressed. We provide a more detailed discussion of this rather counterintuitive aspect of discreteness-noise-driven scatter in \aref{app:Universal_Scatter_Force}.


\subsection{Universality against AMR strategies}
\label{ssec:Universal_AMR_Strategies}

So far, all simulations discussed above adopted our fiducial cell refinement, which causes a cell to be refined whenever it contains more than $N_\text{par/cell} = 4$ particles. This is a fairly aggressive refinement choice compared to most simulations in the literature, which typically adopt $N_\text{par/cell} = 8$  \citep[e.g.][]{Dubois2016MNRAS4633948D, Kim2016ApJ833202K, Read2016MNRAS4592573R, Agertz2020MNRAS4911656A, Park2021MNRAS5086176P, Cadiou2021MNRAS5081189C, Rey2022MNRAS5104208R, Katz2024arXiv241107282K, Storck2025MNRAS539487S}. Unfortunately, it is not really clear what is considered optimal, and different simulations have occasionally used different values, including $N_\text{par/cell} = 4$ \citep{Chiang2024arXiv241103192C}, $5$~\citep{Kravtsov1997ApJS11173K}, $10$~\citep{Ocvirk2008MNRAS3901326O}, $14$~\citep{Alimi2012arXiv12062838A}, or even a redshift-dependent implementation with $N_\text{par/cell} = 2$\textendash$5$ \citep{Klypin2011ApJ740102K, Prada2012MNRAS4233018P}. In particular, it is unclear whether there is a single optimal $N_\text{par/cell}$ to best resolve the tidal evolution of subhalos. Is this choice $\Npar$-dependent? Aside from the increased computational cost, does opting for smaller $N_\text{par/cell}$ always imply better numerical convergence, or can it result in ``over-resolving'' the subhalo, akin to how a softening length in tree-based simulations that is too small causes the system to undergo artificial two-body relaxation?
\begin{figure*}
	\includegraphics[width=0.99\linewidth]{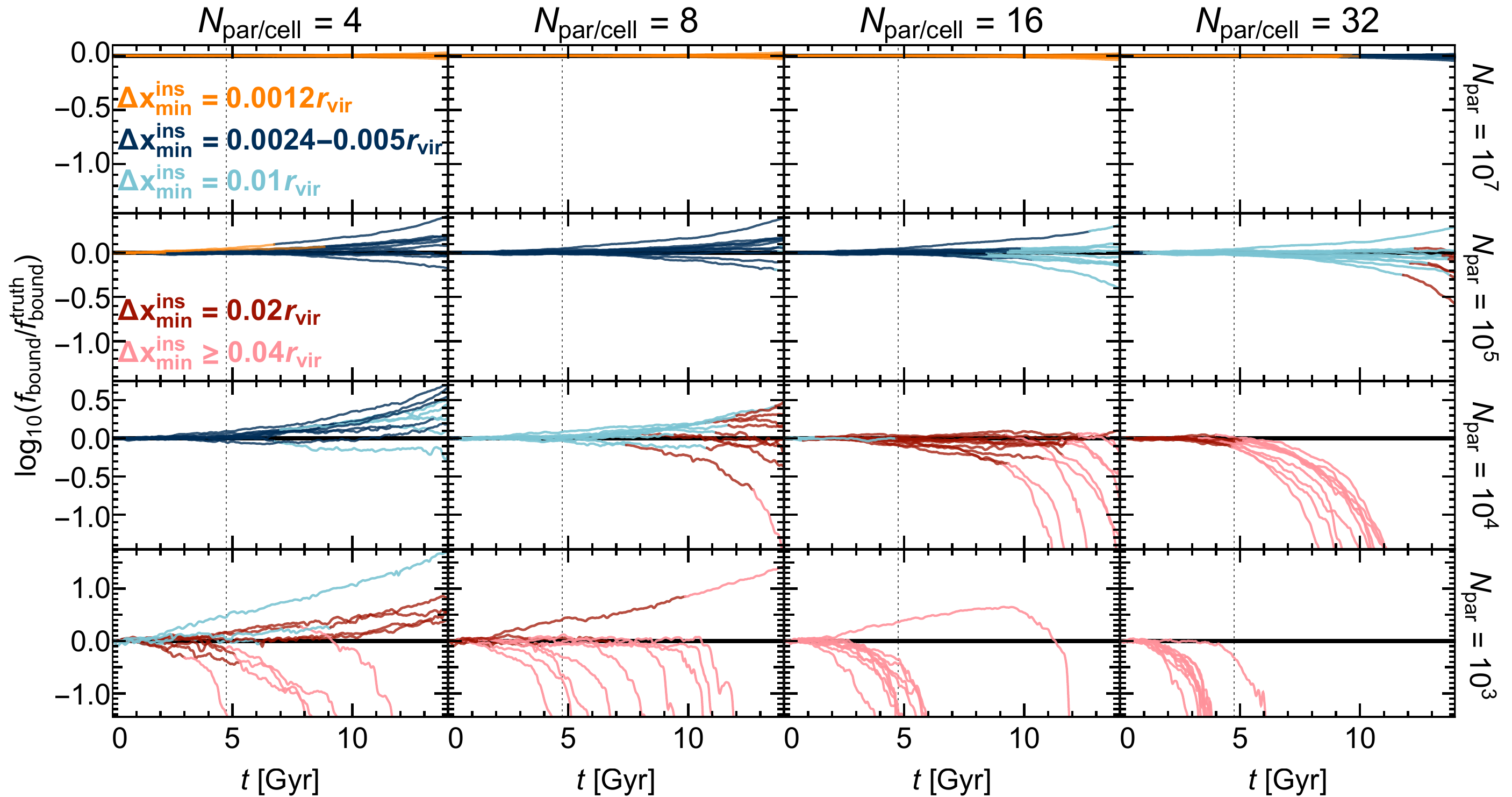}
    \caption{Same as \fref{fig:Force_Conv_MassRes}, but with different columns now showing results for different AMR criteria $N_\text{par/cell} = 4$ (fiducial), $8$, $16$ and $32$ (left to right). All simulation results shown here used a maximum refinement level of 9 for which $\Delta x_\text{min} = 0.0012\rvir$. Note that the specific choice for $N_\text{par/cell}$ has practically no impact when using high mass resolution (e.g. $\Npar = 10^7$). However, for low-$\Npar$ subhalos, larger values of $N_\text{par/cell}$ cause subhalos to become force-unresolved earlier, resulting in enhanced artificial disruption.}
	\label{fig:Force_Conv_Ncell}
\end{figure*}

To address these questions, we repeat the simulations in the leftmost column of \fref{fig:Force_Conv_MassRes} with $\Delta x_\text{min} = 0.0012\rvir$ (i.e., allowing refinement up to level 9) for all four values of $\Npar$, but adopting $N_\text{par/cell} = 8, 16,$ or $32$. The results, together with our fiducial results with $N_\text{par/cell} = 4$, are shown in \fref{fig:Force_Conv_Ncell}. With $\Npar = 10^7$, we find that the highest level of refinement is always achieved, except for the late-time evolution of subhalos evolved under the least stringent refinement condition with $N_\text{par/cell} = 32$. Since $\Delta x^\text{ins}_\text{min} \leq 0.01\rvir$ is never violated, the subhalos are always force-resolved resulting in evolutionary tracks $\fbound(t)$ that are indistinguishable and in perfect agreement with the ``truth,'' despite substantial differences in their AMR grid structures. To illustrate this last point, at $N_\text{par/cell} = 4$, the initial distribution of $10^7$ particles on each refinement level reads: $4.4\%$ (level~5 and below), $13.6\%$ (level~6), $25.7\%$ (level~7), $29.1\%$ (level~8), and $27.2\%$ (level~9), spreading over a total number of $9.8\times 10^7$ cells (force resolution elements). In stark contrast, $N_\text{par/cell} = 32$ yields an initial particle distribution of $24.1\%$ (level~5 and below), $28.0\%$ (level~6), $28.1\%$ (level~7), $15.4\%$ (level~8), and $4.4\%$ (level~9), deposited onto a total of only $6.4\times 10^6$ cells, down by a factor of 15. Even with these drastically different grid structures, a remarkable level of numerical consistency is maintained. 

With $\Npar = 10^5$, $\Delta x^\text{ins}_\text{min}$ noticeably degrades (i.e. increases in size) with increasing $N_\text{par/cell}$. Except for the case with $N_\text{par/cell} = 32$, the subhalos remain ``force-resolved'' (that is $\Delta x^\text{ins}_\text{min} \leq 0.01\rvir$) and the simulations therefore produce convergent evolution on average. At $N_\text{par/cell} = 32$, the overly relaxed refinement choice fails to effectively refine the central regions, resulting in some of the subhalos becoming ``force-unresolved'' at late times, incurring artificially enhanced mass loss. With $\Npar = 10^4$, this trend worsens; subhalos evolved with $N_\text{par/cell} \geq 8$ are all ``force-unresolved'' (well) before $\tH$. With $N_\text{par/cell} \geq 16$, these subhalos experience catastrophic degradation in $\Delta x^\text{ins}_\text{min}$ due to the overly relaxed refinement criterion. Finally, with $\Npar = 10^3$, even our fiducial aggressive refinement with $N_\text{par/cell} = 4$ is unable to properly mass- or force-resolve the subhalos,
and this problem only gets worse with increasing $N_\text{par/cell}$.

Based on these results, we conclude that the force resolution threshold $\Delta x^\text{ins}_\text{min} \leq 0.01\rvir$ holds irrespective of the detailed refinement strategy used. As long as this threshold is factually achieved, the fiducial subhalo used here (which initially has $c=10$ and $\beta = +0.5$) can be considered ``force-resolved.'' A relatively small $N_\text{par/cell}\leq 8$ ($\leq4$) is necessary to ensure that subhalos with $\Npar = 10^5$ ($10^4$) remain ``force-resolved'' down to $\fbound \simeq 0.04$. Smaller $N_\text{par/cell}$ (or larger $\Npar$) will be required to adequately evolve the subhalos to even smaller bound mass fractions. Once satisfied, adopting more aggressive refinement practices (e.g., even smaller $N_\text{par/cell}$ and/or $\Delta x_\text{min}$) is merely a conservative choice at greater computational costs but does not lead to an adverse ``over-resolving'' issue. Lastly, in all cases examined in \fref{fig:Force_Conv_Ncell}, the realization-to-realization scatter in $\fbound$ prior to disruption remains quantitatively unchanged with $N_\text{par/cell}$.


\subsection{Universality against subhalo physical properties}
\label{ssec:Universal_AMR_Force_Sizes}

Having established the robustness of the force resolution threshold against \textit{numerical} mass resolution and refinement choices, we now investigate how the force resolution criterion depends on {\it physical} properties of the subhalos. Thus far, based on the numerical experiments described above, which were all based on an initial subhalo with $c=10$ and $\beta=+0.5$, we inferred a force resolution criterion $\Delta x^{\rm ins}_{\rm min} \leq 0.01 \rvir$. However, we should not expect a more generic force resolution criterion to depend on the virial radius of the subhalo at infall. After all, $\rvir$ is not expected to be of any relevance when it comes to the tidal survival of a subhalo remnant. In addition, for NFW halos in isolation (or subhalos pre-infall), unlike the scale radius $\rsz$, the virial radius does not correspond to any physical feature in the density profile, and its definition in terms of the critical density implies a strong redshift-dependent pseudo-evolution in $\rvir(z)$, even for a static NFW halo without any physical mass accretion \citep{Diemer2013ApJ76625D, Wang2020MNRAS4984450W}. Hence, we seek to reformulate the empirical force resolution criterion used thus far into a more generic, more physical criterion, ideally one that holds independently of physical properties of the subhalo such as its concentration and/or velocity anisotropy. 
\begin{figure*}
	\includegraphics[width=\linewidth]{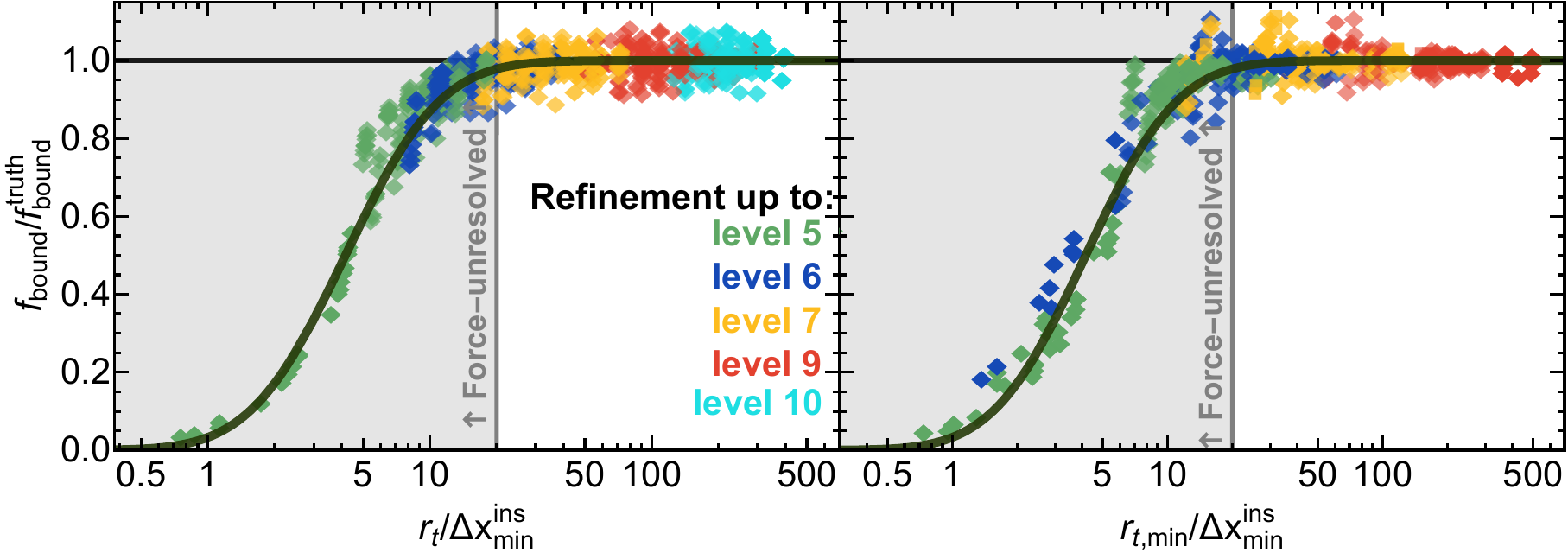}
    \caption{The ratio $\fbound/f^\text{truth}_\text{bound}$ as a function of $\rtmin/\Delta x^\text{ins}_\text{min}$. All results shown here are for simulations with $\Npar = 10^5$. \textit{Left panel:} Results for $(c,\beta)=(10, 0.25)$ subhalos of different initial masses ranging from $10^{-4}\Mhalo$ to $10^{-2.5} \Mhalo$, all evolved along a circular orbit with $\bigRE=0.4$ in the fiducial host halo. For each simulation two points are shown corresponding to the results after one orbital period and after a Hubble time. Different colors indicate simulations run with a different maximum-allowed refinement level, as indicated. \textit{Right panel:} Same as left panel, but for a variety of subhalos of different concentration and velocity anisotropy, and evolved along a wide range of orbits (both circular and eccentric). These results show that subhalos are force-resolved as long as the minimum tidal radius is resolved by at least 20 resolution elements. If this criterion is not satisfied (gray-shaded regions), the subhalo incurs artificially enhanced mass loss. Interestingly, the ratio $\fbound/f^\text{truth}_\text{bound}$ in that case is tightly related to $\rtmin/\Delta x^\text{ins}_\text{min}$ as indicated by the thick black line given by \eref{eqn:fbound_rtid}.}
	\label{fig:Force_Conv_Sizes_rtid}
\end{figure*}

Arguably, the most pertinent physical scale for the tidal evolution of substructure is the tidal radius, in particular, the smallest tidal radius that the subhalo has experienced during its evolution.  Throughout, we adopt the following definition for the tidal radius of \citet[][]{King.62} that accounts for the extended
mass distributions of both host and subhalo and also the centrifugal force from the subhalo's orbital motion\footnote{See \citet{Tollet.etal.17} and \citet{vandenBosch2018MNRAS4743043V} for detailed discussions on different definitions of the tidal radius.}
\begin{align}\label{eqn:Tidal_Radius}
	\rtid = \left[\frac{G m(\rtid)}{\Omega^2 - \frac{\rmd^2\Phi_\rmh}{\rmd r^2}}\right]^{1/3}\,.
\end{align}
 Here, $m(r)$ denotes the instantaneous shell-averaged enclosed mass profile of a subhalo, $\Phi_\rmh(r)$ is the gravitational potential of the host halo, and $\Omega = v/r$ is the angular frequency of the subhalo with $v$ its velocity with respect to the CoM of the host halo. The amount of mass stripping experienced by a subhalo is closely correlated with the {\it minimum} tidal radius 
\begin{equation}\label{eqn:rtmin}
\rtmin \equiv \min[\rtid(t)]\,,
\end{equation}
where $t \in [t_\text{infall}, t_\text{now}]$ spans from the initial instance of subhalo infall to the time of measurement \citep[][]{Jiang.vdBosch.16, Jiang2021MNRAS502621J, Stucker2023MNRAS5214432S, AguirreSantaella2023MNRAS51893A}. In the case of a subhalo on a circular orbit in a static host halo potential, $\rtmin = \rtid(t_\text{now})$. However, more generally, the minimum tidal radius corresponds to that at the most recent peri-centric passage.

In order to investigate whether we can identify a force-resolution criterion in terms of $r_\text{t, min}$, we run a large suite of simulations varying the mass (relative to that of the host halo), concentration, and velocity anisotropy of the initial subhalo, as well as the orbit of the subhalo at infall. For each unique choice of subhalo physical attributes, we simulate ten independent realizations at $\Npar = 10^5$ and one at $\Npar =5\times 10^7$. The results of the latter are interpreted as the ``truth,'' which allows us to quantify the level of numerical convergence for individual subhalos in terms of their $\fbound/f^\text{truth}_\text{bound}$. In particular, we evaluate this ratio at two separate epochs: after one orbital time $t = \torb$, and after a Hubble time $t=\tH$. All of this is repeated for different maximum-refinement levels (i.e., different $\dxmin$), ranging from 5 to 9 with $N_{\rm par/cell}=4$, and once more using $N_{\rm par/cell}=2$ and a maximum refinement level of 10. 

The fiducial parameters for our subhalos are $\msz = 10^{-3} \Mhalo$, $c=10$, and $\beta = +0.25$, and the fiducial orbit is a circular orbit with $\bigRE = 0.4$. We first vary the mass of the subhalo over the range $10^{-4}\Mhalo \leq \msz \leq 10^{-2.5}\Mhalo$ in 0.25~dex increments, keeping all other parameters fixed at their fiducial values, and scaling the virial radius of the subhalos according to $\rvir \propto \msz^{1/3}$ (i.e., all subhalos have the same average density). The left panel of \fref{fig:Force_Conv_Sizes_rtid} plots $\fbound/f^\text{truth}_\text{bound}$ against $\crit$, which indicates the total number of resolution cells with which the instantaneous tidal radius is resolved. Different colors indicate different maximum-refinement levels. For $\crit > 20$ all points lie close to unity, indicating that the bound fractions agree with their ``truth'' value and thus that the results are force-resolved. However, for $\crit < 20$ (gray-shaded region) $\fbound < f^\text{truth}_\text{bound}$. This suggests that in order to be force-resolved the (minimum) tidal radius needs to be resolved by at least 20 resolution elements, i.e.,
\begin{align}\label{eqn:force_res_rule}
\dxins \leq \rtmin / 20\,.
\end{align}
Interestingly, all the results seem to fall along a fairly narrow curve, which is well-fit by 
\begin{align}\label{eqn:fbound_rtid}
\frac{\fbound}{f^\text{truth}_\text{bound}} = 0.5\left\{ \erf \left[ 2.1 \log(0.24 \rtmin/\dxins) \right] + 1 \right\}\,,
\end{align}
indicated by the thick black curve.

In order to test whether this tidal-radius-based criterion is universal, we now run a suite of simulations in which we vary a range of parameters with respect to their fiducial values. More specifically, we vary subhalo concentrations over the physically motivated range $4 \leq c \leq 40$ \citep{Wechsler2002ApJ56852W, Neto:2007vq, Wang2020MNRAS4984450W} and velocity anisotropy over the range $-0.5 \leq \beta \leq +0.5$ \citep{Ludlow2011MNRAS4153895L, Klypin2016MNRAS4574340K, He2024arXiv240714827H}, always keeping the mass fixed at the fiducial value, and we simultaneously vary the initial orbits of the subhalo, including eccentric orbits, covering the ranges $0.2 \leq \bigRE \leq 0.8$ and $e < 0.95$. In addition to these NFW subhalos, we also test initial conditions constructed with Plummer \citep{Plummer1911MNRAS71460P} and Hernquist \citep{Hernquist1990ApJ} density profiles. The results of these simulations are shown in the right panel of \fref{fig:Force_Conv_Sizes_rtid}, with different colors again indicating the different maximum-allowed refinement levels.

As is evident, all of these results fall along the same narrow curve of \eref{eqn:fbound_rtid}, indicating that the force-resolution criterion of \eref{eqn:force_res_rule} is independent of subhalo properties and valid for any orbit. Hence, we conclude that in AMR-based simulations, the tidal evolution of dark matter substructure is properly force-resolved as long as the minimum tidal radius the structure has experienced since infall is resolved with at least 20 resolution cells, irrespective of subhalo mass resolution, refinement details, or physical properties. If violated, the subhalo is force-unresolved, leading to an enhanced mass-loss rate (i.e., a bound mass fraction that is too low) that ultimately results in artificial run-away disruption. 

Note that the fact that force-unresolved systems seem to follow a tight relation between $\fbound/f^\text{truth}_\text{bound}$ and $\rtmin/\dxins$, given by \eref{eqn:fbound_rtid}, suggests that it might be possible to ``correct'' the bound mass fractions of force-unresolved halos in the simulation ex post facto, as long as one is able to accurately infer the value of $\rtmin/\dxins$. We leave it for future work to examine whether this can be used to improve the accuracy of subhalo mass functions extracted from cosmological simulations.


\section{Realization-to-realization Scatter}
\label{sec:Numerical_Convergence}

Dark matter simulations undersample the actual number of dark matter particles by many orders of magnitude, giving rise to discreteness noise that is artificially inflated. Among others, this causes a significant scatter among different random realizations of the same underlying phase-space distribution function. This realization-to-realization scatter also impacts the tidal evolution of subhalos, introducing stochastic errors in the bound mass fractions, even with adequate force resolution \citep{vdBosch.Ogiya.18}. The only way to suppress these errors is to adopt a higher mass resolution, that is, run the simulations using a larger number of particles. 
\begin{figure}
	\includegraphics[width=\linewidth]{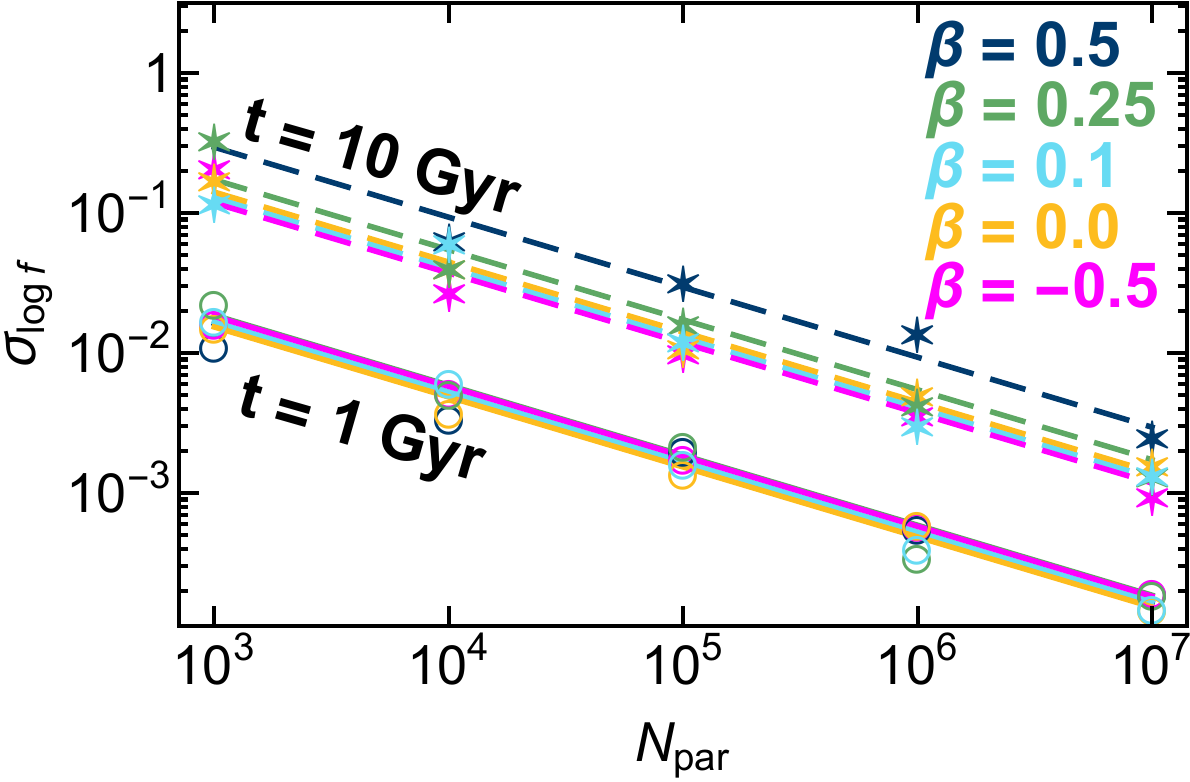}
	\caption{Numeral scatter in the bound mass fractions, $\sigma_{\log f}$, for subhalos of varying $\Npar$ ($x$-axis) and initial velocity anisotropies $\beta$ (color-coded as indicated), evolved along circular orbits with $\bigRE=0.4$. Each data point is measured from ten independent realizations at $t = 1$ (open circle) or $10$~Gyr (star), overlaid with the best-fit scaling $\propto \Npar^{-0.5}$ (solid and dashed lines). Note how at late times $\sigma_{\log f}$ develops a dependence on $\beta$.}
	\label{fig:Sigma_fbound_Scaling_1}
\end{figure}

In this section, we quantify the numerical scatter in $\fbound$ as a function of the initial mass resolution, $\Npar$, of the subhalo at accretion. Previously, \citet{vdBosch.Ogiya.18} studied the tidal evolution of isotropic subhalos evolved on circular orbits and found that $\sigma_{\log f} \propto \Npar^{-1/2} t$, in agreement with simple predictions for discreteness noise. Here, we extend this analysis to anisotropic subhalos with $-0.5 \leq \beta \leq +0.5$, all with our fiducial concentration parameter $c=10$ and focusing only on circular orbits with $\bigRE = 0.4$. As we demonstrate explicitly in Appendix~\ref{app:Universal_Scatter_Force}, our results are independent of the subhalo orbit.
 
The results are shown in \fref{fig:Sigma_fbound_Scaling_1}, where the open circles indicate the scatter in the bound mass fractions inferred after 1 Gyr of evolution in our fiducial host system as a function of the mass resolution. Different colors correspond to different values of $\beta$, as indicated. As is apparent, these results nicely follow the expected $\Npar^{-1/2}$ scaling (colored lines) throughout the range from $\Npar = 10^3$ to $10^7$ and irrespective of the anisotropy of the subhalo. The asterisks show the results inferred after 10~Gyr. Although these still follow the $\Npar^{-1/2}$-scaling for each individual value of $\beta$, the normalization of the $\sigma_{\log f}(\Npar)$ relation is seen to increase with increasing $\beta$. This indicates that the scaling of $\sigma_{\log f}$ with time is not linear. In particular, more radially (tangentially) anisotropic subhalos accumulate realization-to-realization scatter more (less) rapidly. Given that more radially anisotropic subhalos undergo more rapid tidal mass loss, and given that discreteness noise is expected to scale with the square root of $N_{\rm bound} = \fbound \Npar$, we next investigate whether $\fbound$ might be a better indicator of discreteness noise (at a given $\Npar$) than time.

\begin{figure}
	\includegraphics[width=\linewidth]{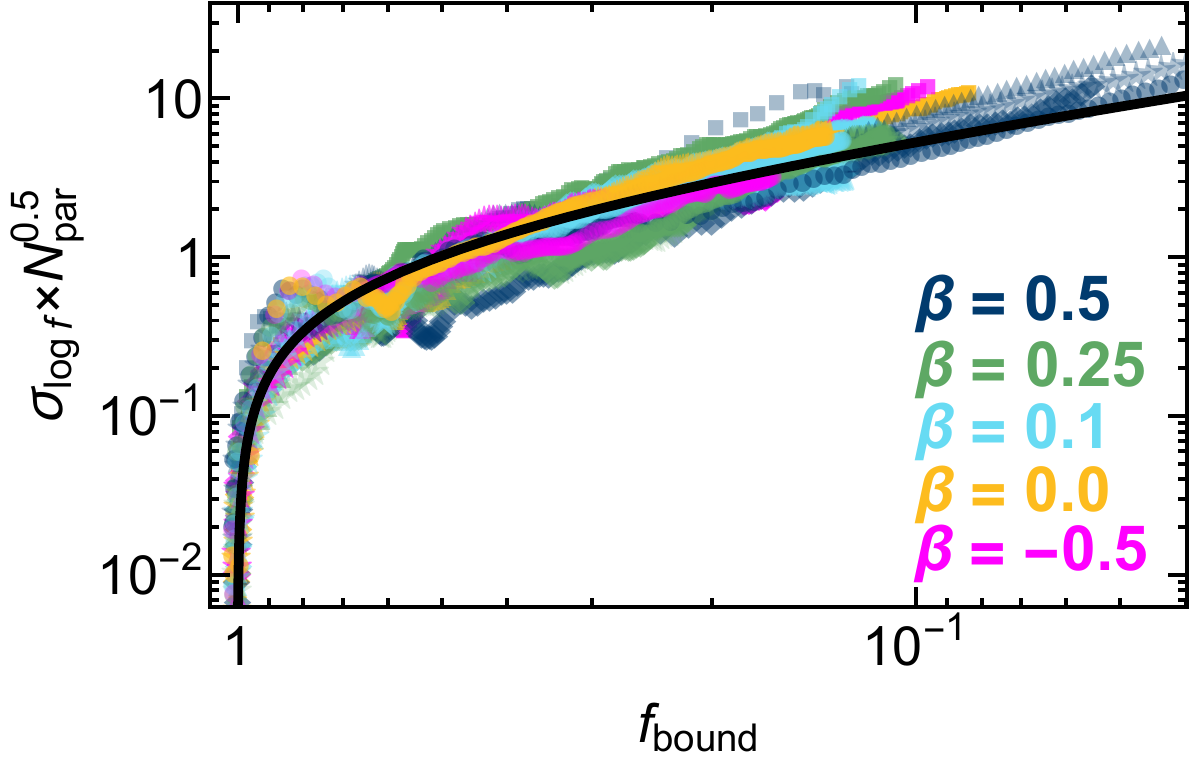}
	\caption{The product of $\sigma_{\log f}$ and $\Npar^{0.5}$ as a function of the instantaneous bound mass fraction, $\fbound$, for the same simulations as used for \fref{fig:Sigma_fbound_Scaling_1}. Each data point indicates the scatter inferred from ten random realizations, plotted against their average value of $\fbound$. Different colors correspond to different values for the initial velocity anisotropy $\beta$, as indicated, while different symbols indicating different mass resolutions; $\Npar = 10^3$ (square), $10^4$ (diamond), $10^5$ (triangle), $10^6$ (star), and $10^7$ (circle). Results are shown over the period from $t=0.01 \Gyr$ to $t=10\Gyr$ or until the epoch when the first subhalo among the ten-realization set disrupts. The thick black line indicates the best-fit relation given by \eref{eqn:sigma_fbound_fit}.}
	\label{fig:Sigma_fbound_Scaling_2}
\end{figure}

\fref{fig:Sigma_fbound_Scaling_2} plots the scatter in $\fbound$ scaled by $\Npar^{0.5}$ as a function of $\fbound$ for all subhalos analyzed in \fref{fig:Sigma_fbound_Scaling_1}. Remarkably, all data points trace a narrow numerical scatter track, irrespective of subhalo anisotropy. This is in stark contrast to the more physical tidal tracks describing the structural evolution of subhalos, which are unique to each anisotropy profile \citep{Chiang2024arXiv241103192C}. The black solid line indicates the best-fit parameterization of this scatter track, which is given by
\begin{align}\label{eqn:sigma_fbound_fit}
\sigma_{\log f} = 1.8\Npar^{-0.5}\fbound^{-0.6}(1-\fbound^{0.5})^{0.8}\,.
\end{align}
We have verified that this expression is universal against variations in the orbital parameters
(see Appendix~\ref{app:Universal_Scatter_Force}) and against variations in the concentrations and density profiles of the subhalos. 

Note that \eref{eqn:sigma_fbound_fit} satisfies the boundary condition $\sigma_{\log f} = 0$ for $\fbound=1$, while it reduces to $\sigma_{\log f} \propto N_\text{bound}^{-0.5} \fbound^{-0.1}$ in the limit of small $\fbound$. Hence, to good approximation the realization-to-realization variance in the bound mass fractions scales as the square root of the instantaneous number of bound particles, in agreement with simple expectations. This universal expression for the stochasticity in the bound mass fractions of subhalos can be used to guard against noisy data in analyzing substructure statistics in cosmological simulations. In particular, as we have demonstrated in \S\ref{ssec:Universal_AMR_Force_Mass}, discreteness noise can lead to spurious survival of subhalos that should have been disrupted or that should have reached values for $\Nbound$ that are below the minimum threshold used to select subhalos in a simulation box. One can avoid a significant contribution of such spurious survivors by removing those subhalos for which \eref{eqn:sigma_fbound_fit} indicates that $\sigma_{\log f}$ exceeds a user-defined threshold value. 
\begin{figure}
        \includegraphics[width=\linewidth]{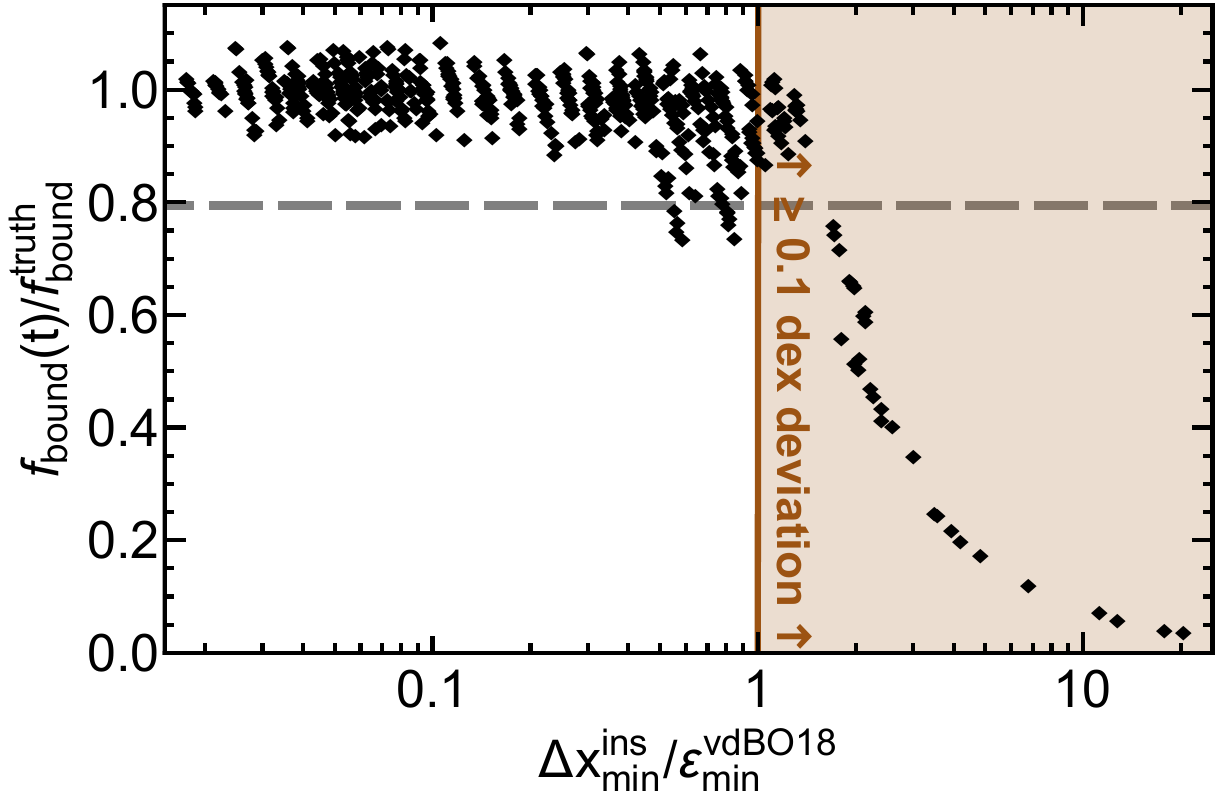}
        \caption{The ratio $\fbound/f^\text{truth}_\text{bound}$ as a function of $\Delta x^\text{ins}_\text{min}/\varepsilon^\text{vdBO18}_\text{min}$, for all data points shown in the left panel of \fref{fig:Force_Conv_Sizes_rtid}. Here, $\varepsilon^\text{vdBO18}_\text{min}$, given by \eref{eqn:vdBO18_Res}, is the critical softening length for tree-based simulations; subhalos are considered force-resolved if simulated with a softening length that is smaller than this minimum required value. Using a larger softening length results in a systematic underestimate of the bound mass fraction that exceeds 0.1~dex (i.e., $\fbound/f^\text{truth}_\text{bound} < 0.794$, indicated by the dashed horizontal line). Note that our AMR-based simulation results fall below this line for $\Delta x^\text{ins}_\text{min}/\varepsilon^\text{vdBO18}_\text{min} > 1$ (brown vertical line), indicating an equivalence between the minimum cell size required in AMR simulations and the minimum softening length required in tree-based simulations.}
	\label{fig:AMR_tree_force}
\end{figure}
%


\section{Implications for Cosmological Simulations}
\label{sec:Current_Literature}


\subsection{AMR- vs. tree-based resolution criteria}
\label{sec:AMR_treecode}

In \S\ref{sec:Universal_AMR_Force} we derived a universal force resolution criterion for properly resolving the tidal evolution of dark matter subhalos using an AMR-based code. In particular, we have demonstrated that a subhalo is force resolved as long as the minimum tidal radius it has experienced is resolved by at least 20 cells. 

We now compare this with a similar criterion, due to \citet[][hereafter vdBO18]{vdBosch.Ogiya.18} but applicable to tree-based simulations. Using a large suite of idealized simulations similar to those used here, vdBO18 inferred that tree-based simulations require an instantaneous Plummer softening length $\varepsilon_\rmP < \varepsilon_{\rm min}^{\rm vdBO18}$, given by
\begin{equation}\label{eqn:vdBO18_Res}
 \varepsilon_{\rm min}^{\rm vdBO18} = 0.56\, \rsz \,\fbound \, g(c) \, \left({r_{\rm half} \over \rsz} \right)^{-1}\,,
\end{equation}
in order for the subhalo to be force-resolved, which vdBO18 defined as having a bound mass fraction that is within 0.1~dex of the converged ``truth'' value. Here, $\rsz$ is the scale radius of the subhalo at accretion, $r_{\rm half}$ is the instantaneous half-mass radius of the subhalo remnant and $g(c) = \ln(1+c) - c/(1+c)$, with $c$ the concentration of the subhalo at accretion. In order to compare this criterion with our force-resolution criterion of \eref{eqn:force_res_rule}, we use the data shown in the left panel of \fref{fig:Force_Conv_Sizes_rtid}. These correspond to simulations of subhalos orbiting along a circular orbit in our fiducial host halo, for which we measured $\fbound$ and $\rtmin$ at two epochs: after one orbital period and after a Hubble time of evolution. At the same epochs, we now compute the half-mass radius of the bound particles, which we use together with the initial scale radius $\rsz$ and the concentration parameter $c$, to calculate $\varepsilon_{\rm min}^{\rm vdBO18}$ using \eref{eqn:vdBO18_Res}. 

\begin{figure}
    \includegraphics[width=\linewidth]{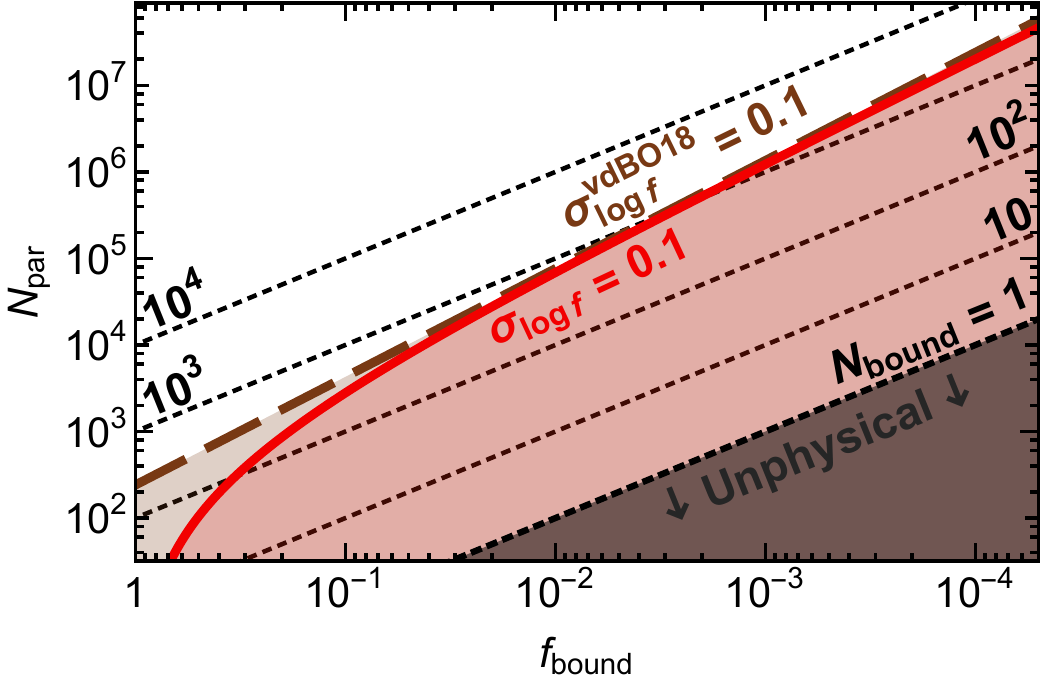}
    \caption{Mass resolution criteria, indicating where discreteness-noise-driven realization-to-realization scatter in the bound mass fractions that exceeds 0.1~dex, based on the work of vdBO18 (brown-dashed line) and our expression for $\sigma_{\log f}$ (red-solid lines). Subhalos that fall below these lines have bound mass fractions that are unreliable due to discreteness noise.  Note the good agreement between these two criteria, which have been inferred from simulations that use different (tree-based vs. AMR-based) architectures for computing the gravitational forces. Black dotted lines indicate fixed numbers of bound particles, and are shown for comparison.} 
	\label{fig:AMR_tree_mass}
\end{figure}

\fref{fig:AMR_tree_force} plots the ratio $\fbound/f_\text{bound}^\text{truth}$ as a function of $\Delta x_{\rm min}^{\rm ins}/\varepsilon_{\rm min}^{\rm vdBO18}$. As is evident, we find that $\fbound/f_\text{bound}^\text{truth} < 0.8$ (i.e., the bound mass fraction is artificially low by more than 0.1 dex, which is considered force-unresolved according to vdBO18) if $\Delta x_{\rm min}^{\rm ins} > \varepsilon_{\rm min}^{\rm vdBO18}$. Interestingly, this is exactly what we would expect if the instantaneous minimum cell size in our AMR simulations plays a similar role as the Plummer softening length in a tree-code as typically suggested \citep[e.g.,][]{OShea2005ApJS1601O}. This implies that we should be able to reformulate our force resolution criterion of \eref{eqn:force_res_rule} as
\begin{equation}\label{eqn:force_res_reform}
\Delta x_{\rm min}^{\rm ins} \leq 0.56 \rsz \,\fbound \, g(c) \, \left({r_{\rm half} \over \rsz} \right)^{-1}\,.
\end{equation}
However, this comes with an important caveat. vdBO18 only tested their force resolution criterion against simulations of isotropic subhalos and predominantly focused on circular orbits (i.e., they only considered one non-circular orbit of moderate eccentricity). Using our much larger suite of simulations, we find that \eref{eqn:force_res_reform} does not adequately describe the force resolution required for highly eccentric orbits and/or for subhalos with strong radial velocity anisotropy. In particular, for our $\beta=0.5$ subhalo, which undergoes tidal-stripping induced core formation, satisfying \eref{eqn:force_res_reform} is not sufficient to guarantee that the subhalo is force resolved. On the other hand, as discussed in \S\ref{ssec:Universal_AMR_Force_Sizes}, the criterion given by \eref{eqn:force_res_rule} remains valid even for these extreme cases (cf. right-hand panel of \fref{fig:Force_Conv_Sizes_rtid}). Hence, we argue that our force resolution criterion based on the minimum tidal radius is more generally applicable and supersedes the criterion presented in vdBO18, and thus that subhalos in tree-based simulations can be deemed force-resolved as long as the minimum tidal radius they have experienced exceeds 20 times the Plummer equivalent softening.
\begin{figure*}
    \includegraphics[width=\linewidth]{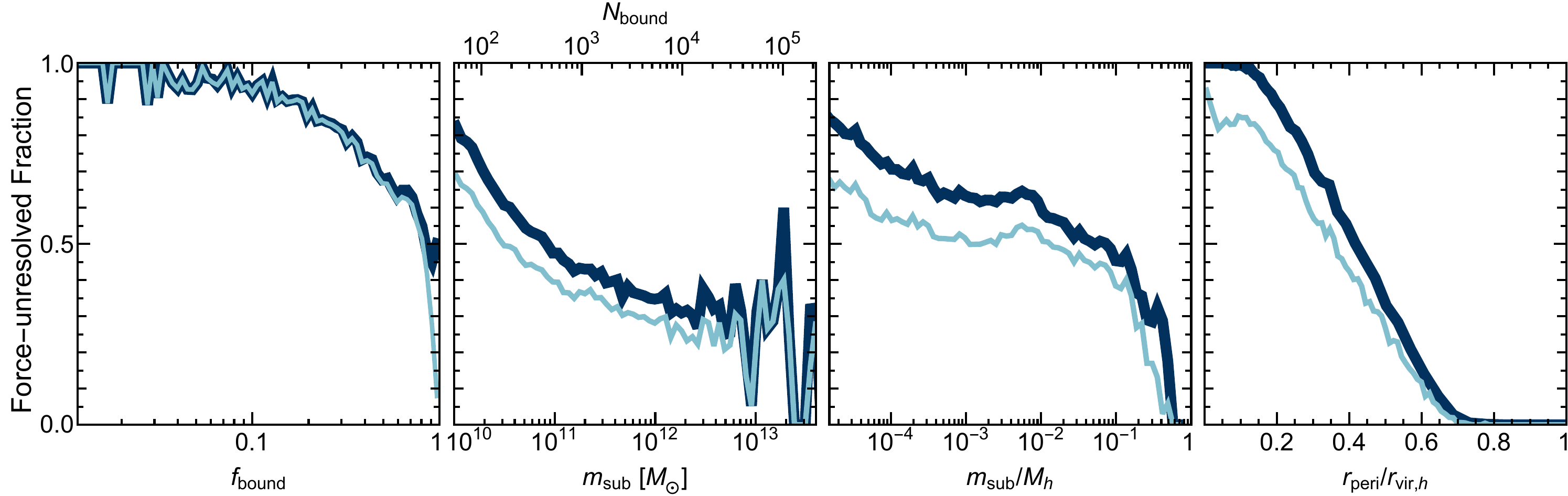}
        \caption{The fraction of Bolshoi subhalos that are force-unresolved according to the universal requirement $r_\text{t,min}/\Delta x^\text{ins}_\text{min} \geq 20$ (dark blue), as a function of $\fbound$, $m_\text{sub}$, $m_\text{sub}/M_\text{h}$, or $r_\text{peri}/\rvirh$ (left to right panels). $M_\text{h}$ and $\rvirh$ denote the mass and virial radius of the subhalo's host halo, respectively. As $r_\text{t,min}$ is computed at peri-center, which can be too conservative for subhalos accreted less than 1~Gyr (about 17$\%$ of the entire sample) and unlikely to have experienced their first peri-centric passage, we additionally show the force-unresolved fractions by assuming that these newly accreted subhalos are all force-resolved (light blue). By either calculation, slightly over half of the entire Bolshoi subhalo sample is force-unresolved. Subhalos are nearly all force-unresolved for $\fbound \lesssim 0.1$ or $\rperi/\rvirh \lta 0.2$. 
        }
	\label{fig:Bolshoi_Force_Add}
\end{figure*}

On the separate issue of mass-resolution-limited artifacts, vdBO18 argued that the bound mass fractions of subhalos are numerically robust against discreteness noise as long as $\sigma_{\log f} \leq 0.1$. Using their suite of idealized tree-based simulations, they find that this translates into the following criterion for the bound mass fraction
\begin{equation}\label{eqn:fbvdB}
 \fbound \geq 0.32 \left({\Npar \over 10^3}\right)^{-0.8}\,, 
\end{equation}
which is indicated as the brown-dashed line in \fref{fig:AMR_tree_mass}. We can compare this directly with our AMR-based results using the universal expression for the numerical scatter $\sigma_{\log f}$ given by \eref{eqn:sigma_fbound_fit}. Setting $\sigma_{\log f}=0.1$, yields the $\fbound(\Npar)$ relation indicated by the red-solid curve, which is in excellent quantitative agreement with the criterion of vdBO18, at least in the heavily stripped limit $\fbound \ll 1$. The disagreement near $\fbound \simeq 1$ is due to the fact that vdBO18 simply extrapolated their findings into this regime, whereas we imposed the physical constraint that $\lim_{\fbound \to 1}\sigma_{\log f} = 0$ when fitting for $\sigma_{\log f}$ as a function of $\fbound$ and $\Npar$. The fact that both criteria are in good agreement indicates that, not surprisingly, the impact of discreteness noise is independent of the simulation architecture. In general, \fref{fig:AMR_tree_mass} shows that it is advisable to exclude subhalos with fewer than a few hundred bound particles from any analysis in order to avoid the results being affected by discreteness noise.


\subsection{Vetting subhalos in cosmological simulations}
\label{sec:Sub_Cosmo_Sim}

An important goal of this work is to provide a unified framework for assessing numerical robustness of simulated subhalos in existing and future cosmological simulations. Here, to obtain some insight, we investigate the fraction of subhalos that satisfies our universal criteria for being force- and/or mass-resolved in a large halo catalog of an existing state-of-the-art cosmological simulation based on AMR. We follow vdBO18 and consider a subhalo mass-resolved if the expected discreteness noise induced scatter $\sigma_{\log f}$ is smaller than $0.1$. Similarly, a subhalo is considered force-resolved if the minimum tidal radius it has experienced exceeds 20 times the minimum resolution element.

As an example, here we consider the Bolshoi simulation \citep{Klypin2011ApJ740102K} performed with the AMR code \textsc{Art} \citep{Kravtsov1997ApJS11173K}. This simulation has $\Delta x_\text{min} = 1.43$~kpc (physical) and a particle mass  $m_\text{DM} = 1.86\times10^8~\Msun$. We use the (sub)halo catalog that has been constructed using the subhalo finder \textsc{Rockstar} \citep[][]{Behroozi2013ApJ762109B}. This catalog is reported to be complete down to a maximum circular velocity of $V_\text{max} = 50 \kms$, which roughly translates to a subhalo mass of $m_\text{sub} \simeq 10^{10}\Msun$, or about 50 particles. We select all first-order non-phantom subhalos\footnote{Phantom is a placeholder object to represent a dark matter subhalo that has been temporarily lost by a halo finder at a given time step; see \cite{Behroozi2013ApJ762109B} and \citet{vandenBosch2017MNRAS468885V} for details.} with $V_\text{max} \geq 50 \kms$ from the Bolshoi halo catalog at redshift $z = 0.0306$.

Consider a subhalo with peak mass $m_\text{peak} \eee \msz = \Npar \times m_\text{DM}$ and instantaneous mass $m_\text{sub} \eee \fbound\times m_\text{peak} = N_\text{bound} \times m_\text{DM}$. To apply the force resolution criterion of \eref{eqn:force_res_rule}, we need to determine the minimum tidal radius $r_\text{r,min}$ of the subhalo. Under the assumption of velocity isotropy, we first compute the density profile of the subhalo bound remnant using the isotropic tidal track of \citet{Green2019MNRAS4902091G} and the concentration of the subhalo at infall. The subhalo catalog also provides the mass and concentration of the host halo, which we use to compute the host halo potential $\Phi_\rmh$ assuming that it follows a spherical NFW profile. Together with the instantaneous position and velocity of the subhalo with respect to its host halo, this allows us to compute both the peri-center of the subhalo orbit, $r_{\rm peri}$ from \eref{apoperi} and the subhalo tidal radius $r_\rmt$ from \eref{eqn:Tidal_Radius}. Throughout we assume that this tidal radius evaluated at $\rperi$ represents the {\it minimum} tidal radius $r_\text{t,min}$, and that $\fbound$ remains comparable between the time of measurement and the most recent peri-center passage. We comment on the limitations of these simplifications in \S\ref{sec:Sub_Discussion}. If we now make the additional optimistic assumption that $\Delta x^\text{ins}_\text{min} = \Delta x_\text{min}$ (i.e., the central region of the subhalo is refined to the maximum allowed level), this allows us to check for each individual subhalo if the condition $\dxins \leq \rtmin / 20$ is satisfied or violated. 

\smallskip

The inferred fractions of force-unresolved first-order subhalos in the Bolshoi catalog are shown as dark blue curves in \fref{fig:Bolshoi_Force_Add} as functions of $\fbound$, $m_\text{sub}$, $m_\text{sub}/\Mhalo$, and $\rperi/\rvirh$ (panels from left to right). For comparison, the light-blue colored lines show the results obtained under the assumption that all subhalos accreted within the last Gyr are force-resolved, as they are unlikely to have experienced their first peri-center passages. As is apparent, this yields force-unresolved fractions that are only marginally smaller. At just over 50\% the overall fraction of subhalos that are deemed force-unresolved based on the criterion of \eref{eqn:force_res_rule} is disturbingly high. The fraction increases with decreasing $\fbound$, $m_\text{sub}$, and $\rperi/\rvirh$. In particular, subhalos with $\fbound \lesssim 0.1$ are all essentially force-unresolved and therefore en route to run-away numerical disruption. Hence, we do not expect to see many long-lived subhalos in this ``transient'' phase. Indeed, only about $2\%$ ($0.02\%$) of all subhalos in the sample have bound mass fractions below $0.1$ ($0.01$). Finally, the monotonic increase in the force-unresolved fraction with decreasing $\rperi/r_\text{h}$ offers a natural explanation for the dearth of subhalos with small halo-centric distances in cosmological simulations.
\begin{figure}
    \includegraphics[width=\linewidth]{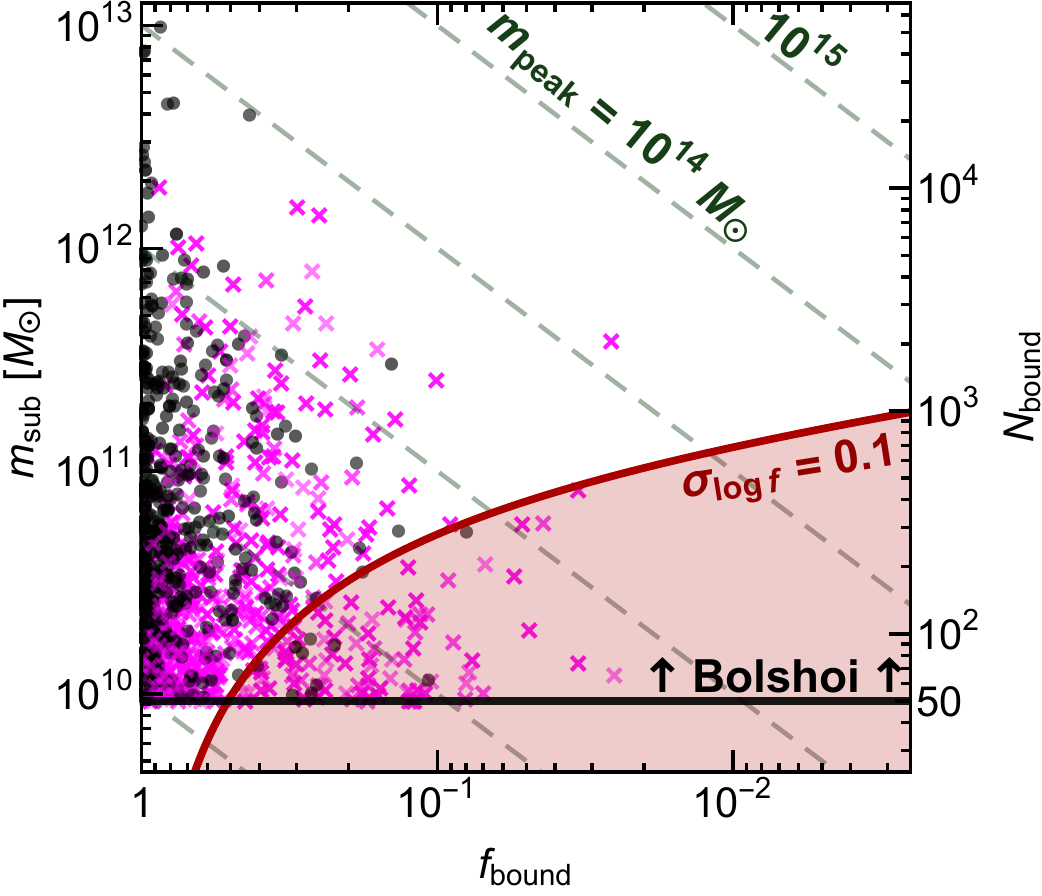}
        \caption{The present-day mass $m_\text{sub}$ against $\fbound$ for a random sample of $10^3$ subhalos in the Bolshoi simulation with $N_{\rm bound} \geq 50$ (horizontal black line). Black dots and magenta crosses indicate subhalos that are considered force-resolved and force-unresolved, respectively. The red-shaded region indicates the part of parameter space where the discreteness-noise-induced scatter in the bound mass fraction $\sigma_{\log f}>0.1$. Subhalos in this region are poorly mass-resolved, giving rise to large uncertainties in their masses. The diagonal dashed lines show lines of constant peak mass, $m_{\rm peak}$, as indicated.}
	\label{fig:Force_Res_Comparison}
\end{figure}

\fref{fig:Force_Res_Comparison} plots the locations of a subset of $10^3$ randomly selected Bolshoi subhalos in the parameter space of $m_\text{sub}$ versus $\fbound$. Subhalos that are deemed force-resolved (unresolved), based on our universal criterion of \eref{eqn:force_res_rule} and the method described above, are indicated as black dots (magenta crosses). For comparison, the solid red curve indicates where the discreteness-noise-induced scatter $\sigma_{\log f} = 0.1$; subhalos below this line therefore have uncertainties in their bound mass fraction exceeding 0.1~dex due to poor mass resolution. The horizontal black line corresponds to $N_{\rm bound}=50$, which roughly corresponds to the cutoff limit $V_{\rm max} = 50\kms$ of the Bolshoi catalog. Note that a large fraction of the subhalos are force-unresolved and/or mass-unresolved. This includes a significant fraction of massive subhalos with $m_{\rm peak} > 10^{12} \Msun$, which are resolved (at accretion) by more than 5,000 particles. 

To make matters worse, in assessing whether a subhalo is force resolved or not, we have assumed that the target resolution scale is always achieved, i.e., $\Delta x^\text{ins}_\text{min} = \Delta x_\text{min}$ in the centers of all subhalos. However, as shown in \fref{fig:Force_Conv_MassRes}, this condition is unlikely to be satisfied for subhalos that are poorly mass-resolved ($\Npar \lesssim 10^3$). Hence, in reality, the fraction of subhalos that is force-unresolved is likely to be even higher than what is shown in Figs.~\ref{fig:Bolshoi_Force_Add} and~\ref{fig:Force_Res_Comparison}. 

Although the above analysis focused only on the \textsc{Rockstar}-based subhalo catalog from the Bolshoi simulation, we suspect the results for other subhalo finders and/or other cosmological simulations to be similar, if not worse. After all, the issues are not really related to specifics of the subhalo finder, and Bolshoi is a typical example of a state-of-the-art cosmological simulation in terms of its mass and force resolution. Furthermore, Bolshoi adopted a redshift-dependent refinement strategy that is fairly aggressive, with $N_\text{par/cell}= 2$\textendash$5$ increasing with decreasing redshifts \citep{Klypin2011ApJ740102K}. As noted in \S\ref{ssec:Universal_AMR_Strategies}, several cosmological simulations have adopted more relaxed refinement choices, using $N_\text{par/cell}=8$ or larger. As illustrated in \fref{fig:Force_Conv_Ncell}, larger $N_\text{par/cell}$ makes it more difficult to reach the target resolution scale, which increases the fraction of subhalos that are force-unresolved.

Given these considerations, we strongly suspect that the masses of many subhalos in cosmological simulations are likely to be unreliable due to inadequate force- and/or mass-resolution. And since the masses of subhalos that are force-unresolved are typically underestimated, this could have a significant impact on the inferred subhalo mass function and other population-level statistics. Note that this problem is in addition to the artificial disruption that artificially suppresses the numbers of subhalos, especially at small halo-centric radii, and which is estimated to impact the subhalo mass functions by as much as 20 percent \citep[][]{Green2021MNRAS5034075G}.


\subsection{Additional complications in a cosmological context}
\label{sec:Sub_Discussion}

The numerical convergence criteria presented in this work are based on idealized $N$-body simulations of subhalo tidal evolution, which by no means fully capture the rich dynamics of structural evolution in a cosmological context. Below, we comment on important complications, potential limitations, and practical challenges in directly translating our results to simulated subhalos in cosmological simulations:

\begin{enumerate}[leftmargin=0.27truecm, labelwidth=0.2truecm]\renewcommand\labelenumi{\bfseries\theenumi}

\item \textbf{Dynamical friction and orbital decay}: For the subhalo-to-host mass ratios $\msz/\Mhalo \ll 0.1$ examined in this work, subhalos experience negligible dynamical friction over a Hubble time \citep[e.g.][]{MBW10}, which justifies our use of a static host halo potential. At a more comparable mass ratio $\msz/\Mhalo \gtrsim 0.1$ where dynamical friction becomes important, subhalo orbital decay in simulations is also affected by the discreteness noise from both the subhalo and the host, which should further increase the numerical realization-to-realization scatter. Indeed, such stochastic dispersion in subhalo orbital decay is present in IllustrisTNG with strong resolution dependence (see Appendix~A of \citet{Lovell2025arXiv250907078L}). This additional source of mass-resolution-dependent scatter is not accounted for in \eref{eqn:sigma_fbound_fit}.
    
\item \textbf{Baryon and feedback physics:} Throughout this work, we have considered the tidal evolution of dark matter subhalos without baryons. Baryonic processes such as adiabatic contraction \citep{Gnedin:2004cx} or supernova feedback \citep{Pontzen.Governato.12} can modify the subhalo density profile pre-infall. In this more complex regime, we expect the tidal-radius-based force convergence criterion to remain valid, as supported by validation tests with subhalos of varying density profiles, physical sizes, and anisotropy-dependent tidal evolution tracks (\fref{fig:Force_Conv_Sizes_rtid}). However, quantifying the additional effect of including gas and associated hydrodynamic processes (e.g., cooling, heating, and feedback) on the stochasticity criteria requires further study.

\item \textbf{Cosmological evolution:} 
Unlike the static and spherical host halo potential assumed in this work, realistic host halo potentials are non-spherical and can evolve rapidly over time. Nevertheless, we suspect that the criteria for force and mass convergence still apply, given their universality against varying subhalo orbits (Figs.~\ref{fig:Force_Conv_Sizes_rtid}, \ref{fig:Subhalo_Anisotropy_fbound_10Gyr}). Regarding stochasticity in tidal evolution, subhalos in a cosmological context also exhibit physical stochasticity arising from complex formation histories and evolution within a large-scale density field. We stress that the numerical scatter quantified in this work, \eref{eqn:sigma_fbound_fit}, accounts only for the \textit{artificial stochasticity} caused by coarse particle down-sampling of subhalo distribution functions, which still persists even if the aforementioned physical scatter is assumed to be absent.

\item \textbf{Accuracy of $\rtmin$ estimates:} Ideally, one would verify force-resolution convergence with on-the-fly measurement of subhalo tidal radii \citep[see][for such an implementation]{Hopkins2023MNRAS5255951H}, which is unfortunately unavailable in current flagship cosmological simulations. With the next-best approach outlined in \S\ref{sec:Sub_Cosmo_Sim}, individual $\rtmin$ is straightforwardly estimated from information readily available in subhalo catalogs, but relies on a few assumptions. First, each $\rperi$ value is inferred from the shell-averaged NFW potential of the host that, in reality, is both time-varying and closer to triaxial \citep{Bailin2005ApJ627647B}. Second, $\rtmin(\rperi)$ is calculated from the subhalo density profile evolved according to the isotropic NFW tidal track, a simplification that fails to account for e.g. velocity anisotropy at infall \citep{Chiang2024arXiv241103192C} or feedback-driven core formation \citep{Pontzen.Governato.12}. We hope to address the uncertainties and possible biases in such $\rtmin$ estimates in future work.
\end{enumerate}


\section{Summary and Conclusions}
\label{sec:Conclusions}

Despite dramatic increases in computing power, properly resolving the substructure of dark matter halos in $N$-body simulations remains subject to numerical artifacts and uncertainties that impact the statistics and demographics of subhalos. In addition to artificial disruption, which causes the abundance of subhalos to be underestimated, an unknown fraction of subhalos are not properly force- and/or mass-resolved, which may adversely impact their properties.
Using idealized simulations of isotropic subhalos, mainly evolved along circular orbits, \citet{vdBosch.Ogiya.18} first reported force- and mass-resolution criteria that are applicable to tree-based simulations with explicit gravitational softening. However, it is unclear whether these resolution criteria are applicable to generic anisotropic subhalos on arbitrary orbits or in AMR-based simulations without explicit gravitational softening.

To this end, we have presented a comprehensive convergence study of the tidal evolution of anisotropic subhalos using the AMR-based simulation code \gamer. The main results are summarized as follows:
\begin{itemize}[leftmargin=0.27truecm, labelwidth=0.2truecm]
    
\item Subhalos are properly force resolved if their instantaneous tidal radius is resolved by at least 20 cells at all times. This universal condition (see Eq.~[\ref{eqn:force_res_rule}]) is both necessary and sufficient, and is agnostic of mass resolution, refinement criteria, or subhalo physical properties like concentration and/or velocity anisotropy. Subhalos that are not properly force resolved experience artificially enhanced mass loss, which ultimately results in artificial disruption. 

\item Because of finite mass resolution, subhalos are subject to discreteness noise that gives rise to realization-to-realization scatter in their bound mass fractions. This scatter depends only on the mass resolution at infall $\Npar$ and the instantaneous bound mass fraction $\fbound$ (see Eq.~[\ref{eqn:sigma_fbound_fit}]). Subhalos that are poorly mass-resolved can either disrupt artificially due to stochastic down-scatter of $\Nbound$, or survive spuriously if $\Nbound$ is up-scattered.

\item Under adequate force resolution, the numerical scatter is unbiased, whereas inadequate force resolution does not inflate the numerical scatter prior to disruption.

\item The minimum (Plummer) softening length required in tree-based simulations to ensure that a subhalo is force-resolved is, to a good approximation, equal to the minimum cell size $\Delta x^\text{ins}_\text{min}$ required in AMR-based simulations, that is, smaller than $(1/20)^{\rm th}$ of the instantaneous tidal radius at all times. Similarly, discreteness-noise-induced numerical artifacts are identical in AMR- and tree-based simulations.

\item Current state-of-the-art AMR-based cosmological simulations do not properly force- and/or mass-resolve subhalos over a sizable portion of the $m_\text{sub}$-$\fbound$ parameter space. As many as 50 percent of all subhalos in a typical subhalo catalog may be affected, even when using fairly conservative quality cuts of $N_{\rm bound} > 100$.

\end{itemize}

Our work has several generic implications for the demographics of subhalos and the satellite galaxies they host. First, the notion that the force resolution criterion relates to the tidal radius offers a natural explanation for one of the most vexing problems related to the demographics of substructure\textemdash the dearth of subhalos in simulations at small halo-centric distances compared to what is inferred from gravitational lensing and/or the demographics of satellite galaxies \citep[e.g.][]{Natarajan2017MNRAS4681962N, Campbel.etal.18, Graus2019MNRAS4884585G, Carlsten2020ApJ902124C}. Since the tidal radius becomes progressively smaller with decreasing distances from the center of the host halo, subhalos with smaller peri-centric distances are more challenging to resolve (see \fref{fig:Bolshoi_Force_Add}). In particular, if the instantaneous tidal radius is inadequately resolved by less than $20 \Delta x_{\rm min}^{\rm ins}$, even if only for a brief period of time during a peri-centric passage, the system becomes force-unresolved and enters a phase with enhanced mass loss. As shown in \citet{vdBosch.Ogiya.18}, this is an irreversible run-away process that, once triggered, inevitably leads to premature disruption. 

Second, discreteness noise due to poor mass resolution can induce large scatter in the bound mass fractions of subhalos, and can even cause the spurious survival of subhalos that would otherwise have disrupted, either physically or artificially. This can make population-level diagnostics (e.g. subhalo mass functions) appear more robust than they really are (see \aref{app:Universal_Scatter_Force}). Hence, any claim that the results of a simulation are reliable simply because population-level statistics have converged should be taken with a grain of salt \citep[see also][]{vdBosch.Ogiya.18}. Rather, claims should be based only on those subhalos that are deemed to be properly force- and mass-resolved. This is possible, albeit potentially tedious, with the universal criteria presented here. Performing convergence validation across subhalo samples partitioned based on these quality flags offers a statistically reliable and numerically efficient approach to identifying suitable problem-specific resolution limits. In fact, the degree of artificially inflated mass loss of subhalos that are force unresolved is closely related to the ratio $r_\text{t,min}/\Delta x^\text{ins}_\text{min}$ through \eref{eqn:fbound_rtid}. This tight correlation suggests that it might be possible to correct the masses of these subhalos for the enhanced mass loss. However, this is only feasible in the absence of significant discreteness noise.

In general, our results are in good agreement with previous studies that examined the survivability and numerical convergence of subhalos in cosmological simulations. \citet{Green2021MNRAS5034075G} found that artificial disruption can reduce the abundance of subhalos by as much as 10 to 20 percent. \cite{Webb2020MNRAS499116W} demonstrated strong resolution dependence in the tidal stripping and disruption of subhalos. \citet{Grand2021MNRAS5074953G} compared a Milky Way-mass halo from the Auriga cosmological zoom-in simulations \citep[][]{Grand2017MNRAS467179G} with a re-simulation at higher mass (by a factor of 64) and force (by a factor of 4) resolutions. This increase in resolution results in a five-fold increase in satellite abundance, one-sixth of which are low-mass subhalos that prematurely disrupted at standard resolution. \citet{Lovell2025arXiv250907078L} reported that in IllustrisTNG, subhalo tidal stripping appears numerically convergent only down to $\fbound = 0.5$ (0.1) for $N_\text{bound} > 10^2$ ($10^3$). These findings, along with our analysis of the Bolshoi subhalos, all paint a consistent picture in which subhalo demographics, especially at the low-mass end and at small halo-centric distances, remain highly susceptible to resolution-limited artifacts in current cosmological simulations.

However, all this is in stark contrast to the recent paper by \citet{He2025ApJ981108H}, who claim that ``current state-of-the-art cosmological simulations have reliably resolved the subhalo population.'' They base this conclusion on a re-analysis of the Aquarius simulations \citep[][]{Springel.etal.08}. In particular, using the subset of simulations run at the second highest resolution level, they study subhalos that have between $10^4$ and $10^5$ particles at accretion. Although a large fraction of these subhalos do not survive until the final snapshot at $z=0$, they claim that their disappearance is due to physical processes related to hierarchical accretion rather than to artificial disruption. While we largely agree with this, their claim that therefore current state-of-the-art cosmological simulations reliable resolve subhalo populations is misleading and incorrect. The simulations that \citet{He2025ApJ981108H} use for their analysis have a Plummer equivalent softening length of $65.8 \pc$, or between $0.003$ to $0.006$ of the $z=0$ scale radius of the host halo. In addition, they focus exclusively on well-resolved subhalos with $10^4 \leq \Npar \leq 10^5$. As shown in \citet{vdBosch.Ogiya.18}, such subhalos resolved with such a small softening length are indeed not expected to undergo artificial disruption unless they come extremely close to the center of the host halo. However, in a typical state-of-the-art cosmological simulation, the vast majority of all subhalos have fewer than $10^4$ particles at accretion and are not resolved with a softening length that small. For comparison, the Bolshoi simulation used here uses a minimum grid size (roughly equivalent to the Plummer equivalent softening length, see \S\ref{sec:AMR_treecode}) of $1.43\kpc$, over 20 times larger than that used in the Aquarius simulations studied by \citet{He2025ApJ981108H}. Furthermore, the work presented here shows that artificial disruption is not the only numerical artifact; a significant fraction of the surviving subhalos are not properly force- and/or mass-resolved, which renders their properties unreliable.

The complications regarding subhalo numerical artifacts addressed here are in addition to and independent of the challenges with subhalo identification. Although there have been many recent efforts to improve the identification of subhalos by incorporating additional information beyond the spatial distribution of particles, there is still little sign of convergence, particularly in the radial distribution of subhalos \citep{Diemer2024MNRAS5333811D, Mansfield2024ApJ970178M, ForouharMoreno2025arXiv250206932F}. Our results indicate that this discrepancy cannot be eliminated by any subhalo finder per se but instead largely originates from the resolution limit intrinsic to each simulation.

Although our work has exposed remaining challenges in properly resolving the tidal evolution of (dark matter) substructure, there are reasons for optimism. As suggested by our force resolution criterion, simulations can be improved by explicitly ensuring that the tidal radii of subhalos remain properly resolved for as long as possible. Although challenging, this can be achieved in both AMR-based and tree-based simulation codes by making the refinement (in the case of AMR) or softening (in the case of tree-codes) adaptive to the local tidal field. Recently, \citet{Hopkins2023MNRAS5255951H} developed such a scheme and showed that it causes a significant boost in the abundance of subhalos, especially at small halo-centric distances. Most importantly, this tidal softening scheme can be implemented without significant computational penalties other than a modest increase in computational cost. Hence, any study whose primary goal it is to resolve the substructure of dark matter halos ought to adopt such a tidal softening or refinement scheme, thereby sidestepping many of the issues with inadequate force resolution and resulting in more reliable predictions for the  abundance and demographics of dark matter subhalos. Unfortunately, the issues with discreteness noise are separate, and their only mitigation is to improve the overall mass resolution of the simulation.

\section*{Acknowledgements}

We acknowledge useful conversations with Philip Mansfield and Kentaro Nagamine. FvdB is supported by the National Science Foundation (NSF) through grants AST-2307280 and AST-2407063. This research is partially supported by the National Science and Technology Council (NSTC) of Taiwan under Grant No. NSTC 111-2628-M-002-005-MY4 and the NTU Academic Research-Career Development Project under Grant No. NTU-CDP-113L7729. We use \texttt{NumPy} \citep{numpy} and \texttt{SciPy} \citep{scipy} for data analysis.


\bibliographystyle{mnras}
\bibliography{MyBibTeX1}

\pagebreak
\appendix
\numberwithin{figure}{section}
\numberwithin{table}{section}
\numberwithin{equation}{section}

\section{Universality of numerical scatter track}
\label{app:Universal_Scatter_Force}

\begin{figure*}
        \centering
	\includegraphics[width=0.5\linewidth]{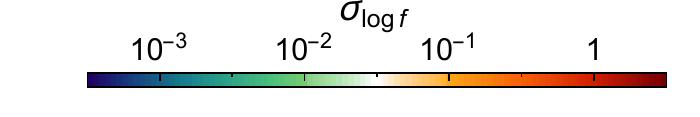}
        \includegraphics[width=\linewidth]{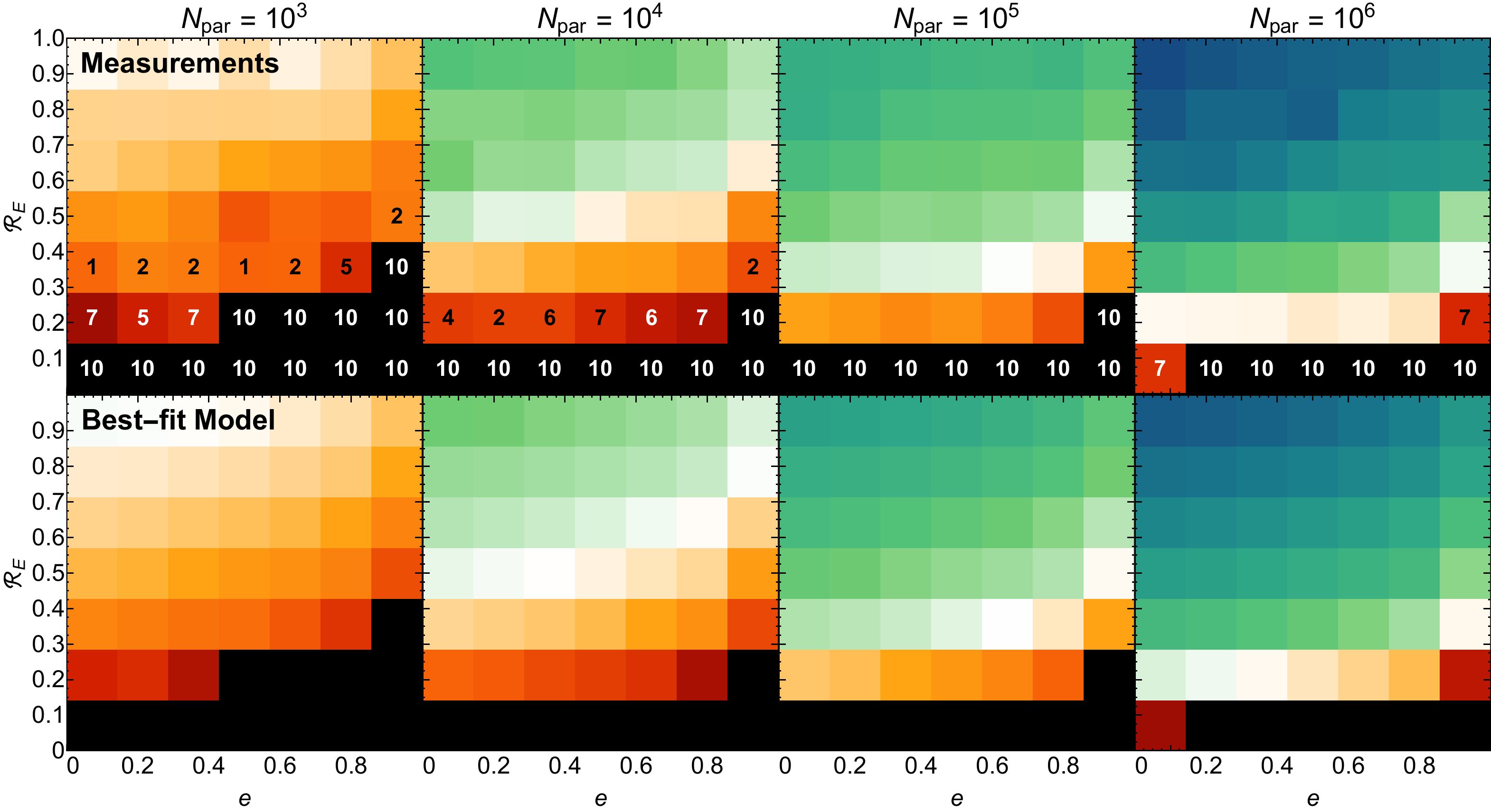}
	\caption{Numerical scatter $\sigma_{\log f}(t=10~\text{Gyr})$ of $\beta = 0.25$ subhalos of $\Npar = 10^3, 10^4, 10^5, 10^6$ (left to right columns), evolved along orbits of different $\bigRE$ and eccentricities $e$. \textit{Top row:} Each pixel is measured from ten independent simulations and annotated with the number of disrupted subhalos, if non-zero; pixels with all ten realizations disrupted are in black. See text for the annotation color and treatment of pixels with partial disruption. \textit{Bottom row:} Model predictions \eref{eqn:sigma_fbound_fit} from the respective $\Npar$ and ten-realization-averaged $\fbound$ for each corresponding pixel in the top row. We observe remarkable agreement between simulation measurements and model predictions. Empirically, we find that the threshold $\sigma_{\log f} \geq 0.3$ is indicative of partial disruption.}
	\label{fig:Subhalo_Anisotropy_fbound_10Gyr}
\end{figure*}

In \S\ref{sec:Numerical_Convergence} we derived an expression (Eq.~[\ref{eqn:sigma_fbound_fit}]) for the discreteness-noise-induced scatter in the bound mass fraction, $\sigma_{\log f}$, as a function of $\Npar$ and $\fbound$. Here we validate the universality of this expression against variations in the subhalo orbital parameters. To that extent, we build upon the simulation suite of \S\ref{ssec:Universal_AMR_Force_Sizes} and conduct additional simulations of a subhalo with $\beta=0.25$ and of fiducial mass and concentration, but evolved over the full range of orbital parameters $\bigRE \in (0,1]$ and $e \in [0,1)$. We consider four different mass resolutions of $\Npar=10^3$, $10^4$, $10^5$ and $10^6$. For each of these, and for each value of $\bigRE$ and $e$ (sampling a $7\times 7$ grid), we run a set of ten random realizations. We use these to compute the realization-to-realization scatter $\sigma_{\log f}$ in the bound mass fractions after $10\Gyr$ of evolution in our fiducial host halo. Each simulation is run with a maximum refinement level such that the subhalo remains adequately force-resolved for at least $10 \Gyr$

The results are indicated in the top row of \fref{fig:Subhalo_Anisotropy_fbound_10Gyr}, with different columns corresponding to simulations with different $\Npar$ as indicated. For each discrete value of $e$ and $\bigRE$, the color coding indicates the value of the scatter, as indicated in the color bar at the top. In addition, we annotate the number of disrupted subhalos in each pixel if non-zero. A pixel is shown in black if all ten realizations are disrupted by $t=10\Gyr$. For each pixel with partial disruption (i.e. at least one of the ten subhalo realizations has been disrupted), we compute $\sigma_{\log f}$ using only the surviving subhalos, which should thus be considered a lower limit to the true variance. For comparison, the bottom row of \fref{fig:Subhalo_Anisotropy_fbound_10Gyr} shows the scatter {\it predicted} based on \eref{eqn:sigma_fbound_fit} for the corresponding $\Npar$ and using the ten-realization-averaged value of $\fbound$ (using $\fbound=0$ for disrupted subhalos). Note the excellent consistency between simulation measurements and model predictions indicating that the subhalo realization-to-realization scatter depends solely on $\Npar$ and $\fbound$ and not on properties of the orbit along which the subhalo has been evolved.

In order to aid in the interpretation of the simulations used to produce \fref{fig:Subhalo_Anisotropy_fbound_10Gyr}, we also run, for each of the $7\times 7$ values of $(\bigRE,e)$, a high-resolution simulation with $\Npar = 5 \times 10^7$. We use the bound mass trajectories $\fbound(t)$ of these simulations as indicative of the ``truth,'' and to compute for any given $\Npar$ the expectation value for the number of bound particles using $\langle \Nbound \rangle = \Npar \, f^\text{truth}_\text{bound}(t=10\Gyr)$. If there were no discreteness-noise-driven scatter (i.e., $\sigma_{\log f} = 0$ and $\fbound = f^\text{truth}_\text{bound}$, as all subhalos are properly force-resolved), then in each given pixel the ten subhalos either all survive (if $\langle \Nbound \rangle > 10$) or all disrupt (if $\langle \Nbound \rangle \leq 10$); recall that throughout we consider a subhalo disrupted if it has 10 or less bound particles remaining. Note, though, that this disruption is
still considered artificial in the sense that the subhalo would have survived if the simulation had been run with higher $\Npar$. We refer to such disruption as mass-resolution-limited disruption, to distinguish it from the force-resolution-limited disruption discussed in \S\ref{sec:Universal_AMR_Force}.

In reality, the tidal evolution of simulated subhalos is subject
to non-zero discreteness-noise-driven scatter, i.e., $\sigma_{\log f} > 0$, which causes $\fbound$ to stochastically deviate from $f^\text{truth}_\text{bound}$. This in turn causes $\Nbound$ to diffuse away from $\langle \Nbound \rangle$, resulting in the following two numerical artifacts. First, if $\langle \Nbound \rangle > 10$ but $\Nbound \leq 10$, then the subhalo should have survived (in the absence of discreteness noise) but instead prematurely disrupted, a process we term \textit{discreteness-noise-induced disruption}. Second, if $\langle \Nbound \rangle \leq 10$ but $\Nbound > 10$, then the subhalo should have undergone mass-resolution-limited disruption but instead spuriously survived, a process we term \textit{spurious survival}.
 
To differentiate between these two stochastic pathways, the respective subhalo disruption counts for pixels in the top panels of \fref{fig:Subhalo_Anisotropy_fbound_10Gyr} are annotated either in black if $\langle \Nbound \rangle \geq 10$ (in absence of discreteness noise, no disruption would have occurred) or in white if $\langle \Nbound \rangle < 10$ (in absence of discreteness noise, all ten subhalos should have undergone mass-resolution-limited disruption). Hence, among the pixels with partial disruption, the black numbers indicate the number of discreteness-noise-induced disruptions, while 10 minus the number in white indicates the number of subhalos that spuriously survived.

Note that the three leftmost pixels at $\bigRE = 0.2$ for $\Npar = 10^3$ indicate that 30 to 50 percent of the subhalos along these orbits spuriously survived. When going to $\Npar = 10^4$, the same pixels now indicate that between 20 to 60 percent of the subhalos experienced discreteness-noise-induced disruption. Despite an order of magnitude increase in mass resolution, the actual number of surviving subhalos has changed very little. This could give rise to a false sense of convergence. In particular, in cosmological simulations low-mass subhalos are more abundant, but are also subject to larger discreteness noise (i.e., larger $\sigma_{\log f}$). Hence, these low-mass subhalos can spuriously survive and artificially ``scatter up'' into higher subhalo mass bins, making the convergence of summary statistics such as subhalo mass functions appear deceptively more optimistic than it really is.

\label{lastpage}
\end{document}